\pdfoutput=1
\documentclass[authoryear,final,3p,times]{elsarticle}
\usepackage{amssymb}
\usepackage{amsthm}
\usepackage{amsmath}
\usepackage{bm}
\usepackage{array}
\usepackage{longtable}
\usepackage{setspace}
\usepackage{multirow}
\usepackage{siunitx}
\usepackage{threeparttable}
\usepackage{verbatim}
\usepackage{tikz}
\usetikzlibrary{positioning, shapes.geometric}
\usetikzlibrary{arrows,shapes,chains}
\usepackage[justification=centering]{caption}
\usepackage{lineno}
\begin{document}
\begin{spacing}{1.0}
\begin{frontmatter}
\title{A critical comparison of the implementation of granular pressure gradient term in Euler-Euler simulation of gas-solid flows}

\author[label1,label2]{Yige Liu}
    \author[label1]{Mingming He}
    \author[label1]{Jianhua Chen}
    \author[label1,label2]{Wen Li}
    \author[label1]{Bidan Zhao}
    \author[label1,label2]{Junwu Wang\corref{cor1}}
    \cortext[cor1]{Corresponding author}
    \ead{jwwang@ipe.ac.cn}
    \address[label1]{State Key Laboratory of Mesoscience and Engineering, Institute of Process Engineering, Chinese Academy of Sciences, Beijing 100190, P. R. China}
    \address[label2]{School of Chemical Engineering, University of Chinese Academy of Sciences, Beijing, 100049, P. R. China}

    \begin{abstract}
Numerical solution of Euler-Euler model using different in-house, open source and commercial software can
generate significantly different results, even when the governing equations and the initial and boundary conditions are
exactly same. Unfortunately, the underlying reasons have not been identified yet.
In this article, three methods for calculating the granular pressure gradient term are presented for two-fluid model of gas-solid flows and implemented implicitly or explicitly into the solver in OpenFOAM$^{\circledR}$:
Method $\uppercase\expandafter{\romannumeral1}$ assumes that the granular pressure gradient is equal to the elastic modulus plus the solid concentration gradient; Method $\uppercase\expandafter{\romannumeral2}$ directly calculates the gradient using a difference scheme; Method $\uppercase\expandafter{\romannumeral3}$, which is proposed in this work, calculates the gradient as the sum of two partial derivatives: one related to the solid volume fraction and the other related to the granular temperature. Obviously, only Methods $\uppercase\expandafter{\romannumeral2}$ and $\uppercase\expandafter{\romannumeral3}$ are consistent with kinetic theory of granular flow.
It was found that the difference between all methods is small for bubbling fluidization. While for circulating fluidization, both methods $\uppercase\expandafter{\romannumeral2}$ and $\uppercase\expandafter{\romannumeral3}$ are capable of capturing non-uniform structures and producing superior results over Method $\uppercase\expandafter{\romannumeral1}$. The contradictory conclusions made from the simulation of different fluidization regime is due to the different contribution of the term related to the granular temperature gradient. Present study concludes that implementation method of granular pressure gradient may have a significant impact on hydrodynamics and is probably a key factor contributing to the observed differences between different simulation software.
    \end{abstract}
\end{frontmatter}

\section{Introduction}
The Euler-Euler model for gas-solid flows that treats both gas and solid phases as interpenetrating continua \citep{anderson1967fluid,gidaspow1994multiphase} has been extensively investigated, where kinetic theory of granular flow (KTGF) is the standard choice for closing the particle phase stress \citep{sinclair1989gas,ding1990bubbling}. This model has gained significant popularity in the simulations of industrial reactors, because of its ability to reconcile economical computational resources and accuracy as well as a clear physics of particle phase stress \citep{wang2020continuum}. The governing equations of Euler-Euler model are a set of nonlinear, strongly coupled partial differential equations, no analytical solutions can be obtained except for extremely simplified situations. Therefore, they have to be solved numerically in practical simulations of gas-solid flows, such as gas-fluidized bed reactors.

Numerical solution of KTGF-based Euler-Euler model has been achieved in many software, such as commercial software Fluent$^{\circledR}$  \citep{fluent2011ansys} and STAR-CCM+$^{\circledR}$  \citep{tandon2014simulation}, in-house software 3D-MFM$^{\circledR}$  \citep{kuipers1992numerical,van2006multiscale} and Saturne$\_$polyphasique@Tlse$^{\circledR}$ \citep{parmentier2008numerical}, open source software MFiX$^{\circledR}$  \citep{syamlal1993mfix}, NEPTUNE$\_$CFD$^{\circledR}$ \citep{neau2020massively} and OpenFOAM$^{\circledR}$ \citep{weller2002derivation,jasak2007openfoam}.
Several recent studies \citep{passalacqua2011ImplementationIterativeSolution,herzog2012comparative,venier2020comparing,nikku2022effect,reyes2022cfd} have found that numerical solutions of two-fluid model using different software options may generate significantly different results, although the governing equations and the initial and boundary conditions are exactly same or very similar. This fact highlights that in addition to physical models, numerical issues are also critical for a successful simulation. Unfortunately, the underlying reasons have not been identified yet.
\citet{passalacqua2011ImplementationIterativeSolution} demonstrated that OpenFOAM$^{\circledR}$  performs closely to MFiX$^{\circledR}$  in predicting the settling suspension, but is significantly different to the results of \citet{parmentier2008numerical} that are obtained using in-house software Saturne$\_$polyphasique@Tlse$^{\circledR}$, when simulating the hydrodynamics of a bubbling fluidized bed.
\citet{herzog2012comparative} compared the two-dimensional results obtained from MFiX$^{\circledR}$  and OpenFOAM$^{\circledR}$, and also examined and discussed those obtained from Fluent$^{\circledR}$. Contrary to OpenFOAM$^{\circledR}$  predictions, computations with MFiX$^{\circledR}$  and Fluent$^{\circledR}$  predicted instantaneous and time-averaged local voidage and velocity profiles which are more comparable with the experimental data existing in the literature. It was concluded that MFiX$^{\circledR}$ and Fluent$^{\circledR}$  were almost equivalently accurate and were technically mature for practical applications, with OpenFOAM$^{\circledR}$  deemed as not mature enough yet.
\citet{venier2020comparing} carried out simulations using both Fluent$^{\circledR}$  and OpenFOAM$^{\circledR}$  to study the characteristics of bubbling and slugging fluidized bed (BFB) with Geldart A, B and D particles. Both software provide precise predictions of the fluidization patterns for Geldart B in a cylindrical bed, whereas Fluent$^{\circledR}$  is less accurate in predicting bubble sizes. However, compared to OpenFOAM$^{\circledR}$, Fluent$^{\circledR}$  exhibited a better comparison with the experimental results when studying Geldart D particles. For Geldart A particles, the results prefer the use of OpenFOAM$^{\circledR}$.
The impact of Fluent$^{\circledR}$ and OpenFOAM$^{\circledR}$  on the simulation outcomes was examined in turbulent and circulating fluidized beds by \citet{nikku2022effect}, despite their differences, both softwares deliver satisfactory results compared with the available measurements. However, one outperforms the other in certain aspects, making it impossible to ascertain which is more accurate for high-velocity fluidized bed modeling. \citet{reyes2022cfd} found that both OpenFOAM$^{\circledR}$  and MFiX$^{\circledR}$  can predict the minimum fluidization velocity in a high-temperature fluidized bed reasonably well, but the wall-to-bed heat transfer coefficient and the fluidization patterns predicted by MFiX$^{\circledR}$  are better than those of OpenFOAM$^{\circledR}$.

In this article, we show that different implementations of the granular pressure gradient term have a minor effect on the hydrodynamics of bubbling fluidized beds, but result in significantly different simulation results of circulating fluidized bed risers, even the governing equations and the initial and boundary conditions are exactly same. The underlying mechanism is also analyzed in detail. Therefore, how to implement the granular pressure gradient term in numerical solutions should be one of the major sources of the observed differences reported in literature.
The remainder of the manuscript is organized as follows. Section 2 the governing equations and constitutive relations of the model are summarized. Section 3 presents the implementation methods of the granular pressure gradient term. Section 4 summarizes the algorithm used to solve the model. Section 5 performs CFD simulations to study the effects of the implementation of the granular pressure term on the simulation results. In Section 6, conclusions are made.

\section{KTGF-based two-fluid model}
In this study, a standard Euler-Euler model is used to simulate the hydrodynamics of gas-solid flows in fluidized beds, in which the gas phase and the solid phase are treated as two interpenetrating continuous fluids. The continuity equations of gas and solid fluids are respectively,
\begin{equation}
    \frac{\partial}{\partial t}\left(\varepsilon_g\rho_g\right)+\nabla\cdot\left(\varepsilon_g\rho_g\bm u_g\right)=0,
    \label{equ:tfm1}
\end{equation}
\begin{equation}
    \frac{\partial}{\partial t}\left(\varepsilon_s\rho_s\right)+\nabla\cdot\left(\varepsilon_s\rho_s\bm u_s\right)=0,
    \label{equ:tfm2}
\end{equation}
where the subscript $g$ denotes gas phase, and $s$ denotes solid phase, $\varepsilon_g$ is the voidage and $\varepsilon_s$ is the solid concentration, $\bm u_g$ is  the gas velocity and $\bm u_s$  is the particle velocity, $\rho_g$ is the gas density and $\rho_s$ is the particle density. The momentum equations are,
\begin{equation}
    \frac{\partial }{\partial t}\left(\varepsilon_g\rho_g\bm u_g\right)+\nabla\cdot\left(\varepsilon_g\rho_g\bm u_g\bm u_g\right)=-\varepsilon_g\nabla p_g+\nabla\cdot\left(\varepsilon_g\bm \tau_g\right)+\varepsilon_g\rho_g\bm g-\bm F_{drag},
    \label{equ:tfm3}
\end{equation}
\begin{equation}
    \frac{\partial }{\partial t}\left(\varepsilon_s\rho_s\bm u_s\right)+\nabla\cdot\left(\varepsilon_s\rho_s\bm u_s\bm u_s\right)=-\varepsilon_s\nabla p_g -\nabla p_s+\nabla\cdot\left(\varepsilon_s\bm \tau_s\right)+\varepsilon_s\rho_s\bm g+\bm F_{drag},
    \label{equ:tfm4}
\end{equation}
where $\tau_g$ is the gas-phase stress tensor and $\tau_s$ is the particle-phase stress tensor, $p_g$ is the shared gas pressure, $p_s$ is the granular pressure, $\bm g$ is the acceleration of gravity, $F_{drag}$ is the interphase drag force. The stress tensors are,
\begin{equation}
    {\bf \tau_g} = \mu_g[\nabla{\bf u}_g+\nabla{\bf u}_g^T]-\frac23\mu_g(\nabla\cdot{\bf u}_g)\bm I,
\label{taug}
\end{equation}
\begin{equation}
    {\bf \tau_s} = \mu_s[\nabla{\bf u}_s+\nabla{\bf u}_s^T]+(\lambda_s-\frac23\mu_s)(\nabla\cdot{\bf u}_s)\bm I,
\label{taus}
\end{equation}
where $\mu_g$ is the shear viscosity of gas and $\mu_s$ is the shear viscosity of particle fluid, $\lambda_s$ is the bulk viscosity of the solid fluid, and $\bm I$ is the unit tensor.

The closure for the drag term is critical for Euler-Euler model. In the present study, when the hydrodynamics of circulating fluidized bed risers are simulated, an energy minimization multiscale (EMMS) drag model \citep{li1994particle} is used, which is able to properly consider the effects of particle clustering structures \citep{lu2009searching}:
\begin{equation}
    {\bf F}_{drag} = \beta_{EMMS}({\bf u_g-u_s})
    \quad
    \text{with}
    \quad
    \beta_{EMMS} = \begin{cases}
        \frac34 C_D\frac{\rho_g\varepsilon_g\varepsilon_s|\bf u_g-u_s|}{d_p}\varepsilon_g^{-2.65}H_d & \varepsilon_s\le 0.35\\
        150\frac{\varepsilon_s^2\mu_g}{\varepsilon_gd_p^2}+1.75\frac{\rho_g\varepsilon_s|\bf u_g-u_s|}{d_p} & \varepsilon_s>0.35
    \end{cases},
    \label{equ::EMMSbeta}
\end{equation}
where
\begin{equation}
    C_D = \begin{cases}
        \frac{24}{Re}(1+0.15Re^{0.687}) & Re<1000\\
        0.44 & Re\ge1000
    \end{cases}
    \quad
    \text{with}
    \quad
    Re = \frac{\varepsilon_g\rho_gd_p|\bf u_g-u_s|}{\mu_g}
\end{equation}
and the drag correction factor $H_d$ that is defined as the ratio of EMMS-predicted drag to the one calculated by the correlation of \citet{1966Mechanics} is
\begin{equation}
H_d = \frac{\beta_{EMMS}}{\beta_{WenYu}}.
\end{equation}
In the simulations of bubbling fluidized bed $\uppercase\expandafter{\romannumeral1}$, the drag model of \citet{syamlal1993mfix} is used:
    \begin{equation}
    {\bf F}_{drag} = \beta({\bf u_g- u_s})
    \quad
    \text{with}
    \quad
    \beta = \frac34 C_D \frac{\rho_g \varepsilon_g \varepsilon_s}{V_r^2 d_p}|\bf u_g - u_s|
    \label{Syamlalbeta}
    \end{equation}
where
    \begin{equation}
    \begin{split}
    C_D = \left( 0.63+4.8\sqrt{\frac{V_r}{Re}} \right)^2
    \quad
    \text{with}
    \quad
    V_r = 0.5[ a-0.06Re+\sqrt{(0.06Re)^2+0.12Re(2b-a)+a^2} ],\\
    a=\varepsilon_g^{4.14},
    \quad
    b=\begin{cases}
    0.8\varepsilon_g^{1.28} & \varepsilon_g \le 0.85 \\
    \varepsilon_g^{2.65} & \varepsilon_g > 0.85
    \end{cases}.
    \end{split}
    \end{equation}
while for the case of the drag model for the simulations of bubbling fluidized bed $\uppercase\expandafter{\romannumeral2}$, the drag model of \citet{gidaspow1994multiphase} is used:
    \begin{equation}
    {\bf F}_{drag} = \beta(\bf u_g-u_s)
    \quad
    \text{with}
    \quad
    \beta = \begin{cases}
        \frac34 C_D\frac{\rho_g\varepsilon_g\varepsilon_s|\bf u_g-u_s|}{d_p}\varepsilon_g^{-2.65} & \varepsilon_s\le 0.2\\
        150\frac{\varepsilon_s^2\mu_g}{\varepsilon_g^2 d_p^2}+1.75\frac{\rho_g\varepsilon_s|\bf u_g-u_s|}{d_p} & \varepsilon_s>0.2
    \end{cases}.
    \label{Gidaspowbeta}
    \end{equation}
The particle phase stress is closed by solving an additional granular energy equation \citep{gidaspow1994multiphase}:
\begin{equation}
    \frac23\left[
\frac{\partial(\varepsilon_s\rho_s\theta_s)}{\partial t}
+
\nabla\cdot(\varepsilon_s\rho_s{\bf u}_s\theta_s)
\right]
-
\nabla\cdot\left(\kappa_s\nabla\theta_s\right)
=
-{p_s{\bf I}}:\nabla{\bf u}_s+{\varepsilon_s\bf \tau_s}:\nabla{\bf u}_s
-\gamma_s\theta_s
-3\beta\theta_s,
\label{equ::granulartemperature}
\end{equation}
where the conductivity of the granular energy $\kappa_s$ is
\begin{equation}
    \begin{split}
          \kappa_s =
\frac{150\rho_sd_p\sqrt{\theta_s\pi}}{384(1+e)g_0}
\left[1+\frac65\varepsilon_s g_0(1+e)\right]^2
+2\rho_s\varepsilon_s^2 d_p(1+e)g_0\sqrt{\frac{\theta_s}{\pi}},
    \end{split}
\end{equation}
and the energy dissipation rate due to inelastic collisions $\gamma_s$ is calculated using
\begin{equation}
    \gamma_s = 12(1-e^2)\varepsilon_s^2\rho_s g_0\frac{1}{d_p}\sqrt{\frac{\theta_s}{\pi}}.
\end{equation}
The granular pressure is,
\begin{equation}
    p_s = \rho_s\varepsilon_s\left[1+2(1+e)\varepsilon_s g_0\right]\theta_s,
\label{equ::ps}
\end{equation}
where $g_0$ is the radial distribution function, which is responsible of enforcing the particle packing limit when the dispersed phase fraction approaches its maximum value. Its common form is,
\begin{equation}
g_0 = [1-(\frac{\varepsilon_s}{\varepsilon_{s,max}})^{\frac{1}{3}}]^{-1}.
\end{equation}
The particle phase shear viscosity is
\begin{equation}
    \begin{split}
            \mu_s=\frac{5}{48}\frac{\rho_s d_p \sqrt{\pi\theta_s}}{\varepsilon_s(1+e)g_0}
        \left[1+\frac45g_0\varepsilon_s(1+e)\right]^2
        +\frac45\varepsilon_s\rho_s d_p g_0(1+e)\sqrt{\frac{\theta_s}{\pi}},
    \end{split}
\end{equation}
and the particle volume viscosity is,
\begin{equation}
    \lambda_s = \frac43\varepsilon_s\rho_s d_p g_0(1+e)\sqrt{\frac{\theta_s}{\pi}}.
\end{equation}

In order to consider the effects of sustained contact between particles and prevent excessive accumulation of particles, it is necessary to set a friction stress model, which plays a role at $\varepsilon_s\ge0.5$. In order to compare with other simulations available in literature, two different frictional stress models are used: when simulating riser flow and bubbling fluidized bed $\uppercase\expandafter{\romannumeral1}$, the frictional pressure is \citep{johnson1987frictional},
\begin{equation}
    p_s^{fr} = \frac{F_r(\varepsilon_s-\varepsilon_{s,min})^{eta}}{(\varepsilon_{s,max}-\varepsilon_s)^p},
\end{equation}
where $F_r=0.05$, $eta = 2$, $p=5$,
$\varepsilon_{s,min}=0.5$, $\varepsilon_{s,max}=0.63$. When simulating bubbling fluidized bed $\uppercase\expandafter{\romannumeral2}$, the frictional pressure is
\citep{schaeffer1987InstabilityEvolutionEquations},
\begin{equation}
p_s^{fr} = 10^{25} (\varepsilon_s - \varepsilon_{s,min})^{10}.
\end{equation}
In both cases, the frictional shear stress is \citep{schaeffer1987InstabilityEvolutionEquations},
\begin{equation}
    \mu_s^{fr} = p_s\frac{\sqrt{2}\sin\phi}{2\sqrt{{\bf S_s:S_s}}}
    \quad
    \text{with}
    \quad
    {\bf S_s} = \nabla{\bf u}_s+\nabla{\bf u}_s^T
    \text{,}
    \quad
    \phi = 30^{\circ}.
\end{equation}

In the simulation of bubbling fluidized beds, the boundary conditions for the granular fluid are set according to the model of \citet{johnson1987frictional}, who developed partial slip boundary conditions for the velocity,
\begin{equation}
\tau_{s,w} = - \frac{\pi}{6}\frac{\varepsilon_s}{\varepsilon_{s,max}}\phi \rho_s g_0 \sqrt{3\theta_s}\bm u_{s,w},
\end{equation}
and for the granular temperature,
\begin{equation}
q_{\theta_s,s} = \frac{\pi}{6} \frac{\alpha_s}{\alpha_{s,max}} \varphi \rho_s g_0 \sqrt{3 \theta_s} |\bm u_{s,w}|^2 - \frac{\pi}{4}\frac{\alpha_s}{\alpha_{s,max}} (1-e^2_{p,w}) \rho_s g_0 \sqrt{3 \theta_s^3},
\end{equation}
where $\tau_{s,w}$ and $q_{s,w}$ are the stress and the granular energy flux at the wall, respectively.

\section{Implementation of granular pressure gradient}
As has been addressed in the Introduction, different methods for implementing the granular pressure gradient term $\nabla p_s$ are available in literature. At earlier studies of gas-solid flow, the granular pressure is assumed to be a function of solid concentration only, therefore, we have \citep{mutsers1977effect,gidaspow1989hydrodynamics,massoudi1992remarks}
\begin{equation}
\nabla p_s = \frac{\partial p_s}{\partial \varepsilon_s} \nabla \varepsilon_s=G(\varepsilon_s) \nabla \varepsilon_s,
\end{equation}
where $G(\varepsilon_s)= \frac{\partial p_s}{\partial \varepsilon_s}$ is the elastic modulus, which is normally closed using various empirical correlations derived from experimental measurements \citep{wang2020continuum}. In modern KTGF-based Euler-Euler model, this assumption is inherited, which results in the following widely used (but incomplete) implementation method.

\subsection{Method $\uppercase\expandafter{\romannumeral1}$}
With the above-mentioned assumption, method $\uppercase\expandafter{\romannumeral1}$ for implementing the granular pressure gradient term in the KTGF-based Euler-Euler model is \citep{weller2002derivation,passalacqua2011ImplementationIterativeSolution,liu2014cfd,venier2014development,venier2016numerical,nikku2019comparison}
\begin{equation}
\nabla p_s = \frac{\partial p_s}{\partial \varepsilon_s} \nabla \varepsilon_s,
\end{equation}
where the elastic modulus is not an empirical correlation anymore, but is derived from KTGF, when simulating the hydrodynamics of circulating fluidized bed risers and bubbling fluidised beds $\uppercase\expandafter{\romannumeral1}$:
\begin{equation}
\begin{split}
\frac{\partial p_s}{\partial \varepsilon_s} = \rho_s \left[1+\varepsilon_s(1+e)\left(4g_0+2\frac{\partial g_0}{\partial \varepsilon_s} \varepsilon_s \right)\right]\theta_s + Fr\frac{ eta(\varepsilon_s - \varepsilon_{s,min})^{eta-1}(\varepsilon_{s,max}-\varepsilon_s)+p(\varepsilon_s-\varepsilon_{s,min})^{eta} }{(\varepsilon_{s,max}-\varepsilon_s)^{p+1}},
\label{Ges}
\end{split}
\end{equation}
when simulating the hydrodynamics of bubbling fluidised beds $\uppercase\expandafter{\romannumeral2}$, it is
\begin{equation}
\begin{split}
\frac{\partial p_s}{\partial \varepsilon_s} = \rho_s\left[1+\varepsilon_s(1+e)\left(4g_0+2\frac{\partial g_0}{\partial \varepsilon_s} \varepsilon_s \right)\right]\theta_s + 10^{26}(\varepsilon_s - \varepsilon_{s,min})^9,
\label{Ges2}
\end{split}
\end{equation}
In both expressions, the partial derivation of radial distribution function is
\begin{equation}
\frac{\partial g_0}{\partial \varepsilon_s}= \frac{1}{ 3 \varepsilon_{s,max}[ (\frac{\varepsilon_s}{\varepsilon_{s,max}})^{\frac{1}{3}} - (\frac{\varepsilon_s}{\varepsilon_{s,max}})^{\frac{2}{3}} ]^2 }.
\end{equation}
This implementation is based on the assumption that the granular pressure is only a function of solid concentration, which is incomplete indeed according to Eq.(\ref{equ::ps}). The granular pressure is not only a function of solid concentration $\varepsilon_s$ but also a function of granular temperature $\theta_s$. In view of this critical deficiency, there cames another implementation method.

\subsection{Method $\uppercase\expandafter{\romannumeral2}$}
The discretization scheme (i.e. the Gauss linear scheme) can be involved directly to compute the granular pressure gradient \citep{syamlal1993mfix,LI2014170,cai2022transition,wartha2022importance}:
\begin{equation}
\nabla p_s = \text{grad}(p_s).
\end{equation}
This discretization results in the cell-centered value of $\nabla p_s$, when constructing the gas pressure Poisson equation and the solid-phase continuity equation (see Section 4 for details), the face-centered value $(\nabla p_s)_f$ needs to be obtained by interpolation. However, the checkerboard effect of granular pressure $p_s$ that is similar to the gas pressure field $p_g$ would occur (such a problem does not exist when staggered variable arrangements  is used, for example {MFiX}$^\circledR$). So present implementation method should be consistent with the Rhie-Chow interpolation to overcome the pressure checkerboard problem:
\begin{equation}
\nabla p_s = \text{snGrad}(p_s) = \nabla^{\perp}p_s = \frac{p_{s,P} - p_{s,N}}{|\bm d_P - \bm d_N|},
\end{equation}
where the face-centered $\nabla^{\perp}p_s$ is directly calculated by its neighbour cells P and N. It should be noted that linearization of $\nabla p_s$ is needed to significantly improve the convergence since $\nabla p_s$ tends to be infinite when the solid concentration is approaching to $\varepsilon_{s,max}$. Therefore, we have
\begin{equation}
\begin{split}
\nabla^{\perp}p_s = min(\text{snGrad}( p_s), k),
\end{split}
\end{equation}
where $k$ is an empirical constant.

\subsection{Method $\uppercase\expandafter{\romannumeral3}$}
In view of the physical nature of granular pressure, present study proposes the third method for implementing the granular pressure gradient term:
\begin{equation}
\nabla p_s = \frac{\partial p_s}{\partial \varepsilon_s} \nabla \varepsilon_s + \frac{\partial p_s}{\partial \theta_s} \nabla \theta_s
\label{psestheta}
\end{equation}
where the first term $\frac{\partial p_s}{\partial \varepsilon_s} \nabla \varepsilon_s$ is calculated as in the method $\uppercase\expandafter{\romannumeral1}$, and the second term considers the effects of granular temperature gradient. Its specific expression can be obtained according to the selected model (Eq. (\ref{equ::ps})),
\begin{equation}
\frac{\partial p_s}{\partial \theta_s} = \varepsilon_s \rho_s [ 1+2(1+e)\varepsilon_sg_0(\varepsilon_s) ].
\label{equ::pstheta}
\end{equation}
Note that since the used empirical models for the frictional granular pressure are not a function of granular temperature, this term only contains the kinetic and collisional contributions to $\frac{\partial p_s}{\partial \theta_s}$.

In summary, the three methods for implementing $\nabla p_s$ and the implicit and explicit implementation of each method that will be presented in Section 4.3 result in six implementation methods in total, and comparison of their effects on the simulation results is the main concern of present study.

\section{Numerical algorithm}
Present study uses OpenFOAM$^\circledR$ to solve the Euler-Euler model using collocated grids, and the numerical algorithm summarized below is mainly based on the works of \citet{passalacqua2011ImplementationIterativeSolution} and \citet{venier2016numerical}. On the other hand, a major difference between present study and their study is the method for handling the interphase drag force term (semi-implicit method vs fully implicit method).

\subsection{Discretized momentum equations}
The phasic momentum equations are treated in fully conservative form \citep{passalacqua2011ImplementationIterativeSolution,venier2014development,li2017simulation}. The semi-discrete form of the momentum equations Eqs. (\ref{equ:tfm3}), (\ref{equ:tfm4}) are, respectively
\begin{equation}
\mathbb A_g \bm u_g = \bm b_g - \varepsilon_g \frac{\nabla p_g}{\rho_g} + \varepsilon_g \bm g + \frac{\beta}{\rho_g} \bm u_s,
\end{equation}
\begin{equation}
\mathbb A_s \bm u_s = \bm b_s - \varepsilon_s \frac{\nabla p_g}{\rho_s}- \frac{\nabla p_s}{\rho_s} + \varepsilon_s \bm g + \frac{\beta}{\rho_s} \bm u_g.
\end{equation}
The matrices $\mathbb A_s$ and $\mathbb A_g$, and the corresponding vectors of source terms $\bm b_s$, $\bm b_g$ include the contributions of the transient term, the convection term, the diffusion term and the implicit drag term.
The interaction between two phases (the drag term) is treated with a semi-implicit method \citep{rusche2003computational}, and the explicit part of the drag term is included in the source term. In order to accelerate the solution, two matrices $\mathbb D_k$ and $\mathbb H_k$ are defined,
\begin{equation}
\mathbb D_k = diag(\mathbb A_k),
\end{equation}
\begin{equation}
\mathbb H_k = \bm b_k - (\mathbb A_k - \mathbb D_k)\bm u_k^{pre},
\end{equation}
where the subscript $k=g,s$. The matrix $\mathbb D_k$ contains the diagonal coefficients of the velocity matrix, and $\mathbb H_k$ represents the off-diagonal part of the velocity matrix and the source part. $\bm u_k^{pre}$ is the pre-existing value of the velocity. So the momentum equations in semi-discrete form become,
\begin{equation}
\mathbb D_g \bm u_g = \mathbb H_g - \varepsilon_g \frac{\nabla p_g}{\rho_g} + \varepsilon_g \bm g + \frac{\beta}{\rho_g} \bm u_s,
\label{equ::dismomeng1}
\end{equation}
\begin{equation}
\mathbb D_s \bm u_s = \mathbb H_s - \varepsilon_s \frac{\nabla p_g}{\rho_s}- \frac{\nabla p_s}{\rho_s} + \varepsilon_s \bm g + \frac{\beta}{\rho_s} \bm u_g.
\label{equ::dismomens1}
\end{equation}
Dividing Eqs. (\ref{equ::dismomeng1}),(\ref{equ::dismomens1}) by $\mathbb D_k$ to get,
\begin{equation}
\bm u_g = \bm u_g^* + \frac{1}{\mathbb D_g}\left( - \varepsilon_g \frac{\nabla p_g}{\rho_g} + \varepsilon_g \bm g + \frac{\beta}{\rho_g} \bm u_s \right),
\label{equ::dismomeng}
\end{equation}
\begin{equation}
\bm u_s = \bm u_s^* + \frac{1}{\mathbb D_s}\left( - \varepsilon_s \frac{\nabla p_g}{\rho_s}- \frac{\nabla p_s}{\rho_s} + \varepsilon_s \bm g + \frac{\beta}{\rho_s} \bm u_g \right).
\label{equ::dismomens}
\end{equation}
$\bm u_k^*$ is the pseudo-velocity of $k$-phase and satisfies,
\begin{equation}
\bm u_k^* = \frac{\mathbb H_k}{\mathbb D_k}.
\label{u*}
\end{equation}
It can be seen that the velocity values are actually obtained by explicitly treating the influence of the surrounding grids and solving simple linear algebraic equations, thus avoiding the iterative solution of the nonlinear momentum equations and therefore saving the computational cost but reducing the stability. Moreover, we emphasize that the granular pressure gradient term $\frac{\nabla p_s}{\mathbb D_s\rho_s}$ is always treated explicitly in the solution of solid-phase momentum conservation equation, no matter how it is treated in the solid-phase continuity equation presented in Section 4.4. Furthermore, the expression for $\nabla p_s$ in the solid-phase momentum conservation equation is always consistent with the selected method for implementing $\nabla p_s$ in the solid-phase continuity equation. Finally, the interphase drag term is handled semi-implicitly in present study, which is sufficient for the relatively coarse particles studied here \citep{venier2016numerical}, since the interphase coupling is moderately strong. However, if the interphase coupling is very strong, this term should be handled fully implicitly using for instance the partial elimination algorithm \citep{spalding1981numerical,passalacqua2011ImplementationIterativeSolution,venier2016numerical}.

\subsection{Gas pressure Poisson equation, interpolation and reconstruction practices}
When the finite volume method is used to solve the momentum equation and the continuity equation in the arrangement of collocated grids, the checkerboard problem occurs \citep{Patankar1980numerical,ferziger2002computational}, due to the fact that the surface velocity is related to the pressure of alternating grids, not to the pressure of consecutive grids. In order to avoid this phenomenon, staggered grids can be used to store the velocity and scalar variables (gas pressure, voidage, solid concentration and granular temperature) in different grid systems. However in this manner, for two-dimensional and three-dimensional cases, three sets and four sets of staggered grids are needed to store the velocity components and scalar variables, and the memory cost is huge. Besides, for non-Cartesian grids, staggered grid systems are difficult to construct, especially for unstructured grids \citep{hsu1982curvilinear,peric1985finite,demirdzic1982finite,darwish2016finite}.
In this study, the collocated grid is used, the variables are all stored at the cell centers. In this case, the discretized momentum equation is usually used for interpolation calculation, the interface flux is interpolated by the cell-centered values on both sides, which makes it impossible to perceive the checkerboard non-uniform field. Therefore,  a special momentum interpolation method is needed to overcome the checkerboard problem \citep{pascau2011cell,zhang2014generalized,bartholomew2018unified}, such as the well-known Rhie-Chow interpolation \citep{rhie1983numerical}. 

Since the densities of gas and solid are constants and $\varepsilon_{s}+\varepsilon_{g}\equiv1$, it is easy to derive the volumetric conservation equation from the continuum equations of gas and solid phases,
    \begin{equation}
    \nabla \cdot \bm u = \nabla \cdot (\varepsilon_{s} \bm u_s + \varepsilon_{g} \bm u_g ) = 0,
    \label{u}
    \end{equation}
after the discretization using finite volume method and the Gauss divergence theorem, the discretized volumetric conservation equation becomes,
    \begin{equation}
    \sum_f \varphi = \sum_f [(\varepsilon_{s})_f \varphi_s + (\varepsilon_{g})_f \varphi_g ] = 0,
    \label{phi}
    \end{equation}
where $\sum_f$ denotes the face sum corresponding to the current mesh, $\sum_f \varphi$ is the total volumetric flux, $\varphi_s$ and $\varphi_g$ are velocity fluxes stored at a face center ($\varphi_k = (\bm u_k)_f \cdot \bm S$), which are scalars and are obtained by calculating the dot-product of velocities according to Eq. (\ref{equ::dismomeng}) and (\ref{equ::dismomens}) with the surface area vector $\bm S$:
    \begin{equation}
    \varphi_s = \varphi_s^*  -\left(\frac{1}{\rho_s \mathbb D_s}\right)_f \nabla^{\perp} p_s |\bm S| -\left(\frac{\varepsilon_s}{\rho_s \mathbb D_s}\right)_f \nabla^{\perp} p_g |\bm S| + \left(\frac{\beta}{\rho_s \mathbb D_s}\right)_f (\bm u_g)_f \cdot \bm S + \left(\frac{\varepsilon_{s}}{\mathbb D_s}\right)_f \bm g \cdot \bm S,
    \label{phis}
    \end{equation}
    \begin{equation}
    \varphi_g = \varphi_g^* -\left(\frac{\varepsilon_g}{\rho_g \mathbb D_g}\right)_f \nabla^{\perp} p_g |\bm S| + \left(\frac{\beta}{\rho_g \mathbb D_g}\right)_f (\bm u_s)_f \cdot \bm S + \left(\frac{\varepsilon_{g}}{\mathbb D_g}\right)_f \bm g \cdot \bm S,
    \label{phig}
    \end{equation}
where $\varphi_k^*=(\bm u_k^*)_f \cdot \bm S$ is the $k$-phase pseudo-velocity at the face center interpolated through neighbouring cell-centered pseudo-velocity $\bm u^*_k$, $\nabla^{\perp}$ denotes the surface-normal gradient, $(...)_f$ denotes the linear interpolation operator from cell-center to face-center. Hence, we have for example
    \begin{equation}
    (\bm u_k)_f = \omega \bm u_{k,P} + (1-\omega) \bm u_{k,N},
    \label{interpolate}
    \end{equation}
    \begin{equation}
(\varepsilon_s)_f=\omega \varepsilon_{s,P}+(1-\omega) \varepsilon_{s,N},
\end{equation}
    where the subscript $P$ denotes the current cell and the subscript $N$ denotes the neighbor cell that shares the same face with the current cell. $\omega = \frac{\overline{fN}}{\overline{fN}+\overline{fP}}$ is the distance weighted coefficient, where $\overline{fN}$ and $\overline{fP}$ are respectively the distances between the center of cell $N$ and the face center $I$ and between the center of cell $P$ and the face center $I$.
    It is worth noting that the term $p_g$ in Eq. (\ref{phis}), (\ref{phig}) is stored at the cell center and its corresponding surface-normal gradient term $\nabla^{\perp} p_g = \frac{p_{g,P} - p_{g,N}}{|\bm d_P - \bm d_N|}$ is stored at the face center shared by cell $P$ and cell $N$, where $\bm d$ is the cell-centered coordinate vector. This term is implemented according to \citet{rhie1983numerical} and \citet{peric1988comparison} to substitute the gradient term calculated by $(\nabla p_g)_f$, therefore, the pressure of the current cell is only discretized and calculated using the adjacent cells. As a result, the final matrix is a compact arrangement, avoiding the checkerboard pressure field and enhancing the convergence of numerical solution.

    Substituting Eq. (\ref{phis}), (\ref{phig}) into Eq. (\ref{phi}), we can get the governing equation for updating the gas pressure field,
    \begin{equation}
    \sum_f \left \{
                \left[
        (\varepsilon_{s})_f\left(\frac{\varepsilon_s}{\mathbb D_s \rho_s}\right)_f + (\varepsilon_{g})_f \left(\frac{\varepsilon_g}{\mathbb D_g \rho_g}\right)_f
    \right]
    |\bm S|\nabla^{\perp} p_g
        \right\}
    = \sum_f \varphi^0,
    \label{Poisson}
    \end{equation}

    where $\varphi^0$ is the total volume flux without the contribution of the gas pressure gradient term, it is expressed as,
    \begin{equation}
    \varphi^0 =  (\varepsilon_{s})_f \varphi^0_s + (\varepsilon_{g})_f \varphi^0_g,
    \label{phi0}
    \end{equation}

    with
    \begin{equation}
    \varphi^0_s =\varphi_s^*  -\left(\frac{1}{\rho_s \mathbb D_s}\right)_f \nabla^{\perp} p_s |\bm S| + \left(\frac{\beta}{\rho_s \mathbb D_s}\right)_f (\bm u_g)_f \cdot \bm S + \left(\frac{\varepsilon_{s}}{\mathbb D_s}\right)_f \bm g \cdot \bm S,
    \label{phic0}
    \end{equation}
    \begin{equation}
    \varphi^0_g = \varphi_g^* + \left(\frac{\beta}{\rho_g \mathbb D_g}\right)_f (\bm u_s)_f \cdot \bm S + \left(\frac{\varepsilon_{g}}{\mathbb D_g}\right)_f \bm g \cdot \bm S.
    \label{phif0}
    \end{equation}

    Once the gas pressure equation Eq. (\ref{Poisson}) is solved, the new gas pressure field stored at the cell center can be obtained and is used to correct the velocity flux of each phase stored at the face center  \citep{weller2002derivation,rusche2003computational}:
    \begin{equation}
    \varphi_s = \varphi_s^0 - \left(\frac{\varepsilon_s}{\mathbb D_s\rho_s}\right)_f \nabla^{\perp} p_g |\bm S|,
    \label{newphis}
    \end{equation}
    \begin{equation}
    \varphi_g = \varphi_g^0 - \left(\frac{\varepsilon_g}{\mathbb D_g \rho_g}\right)_f \nabla^{\perp} p_g |\bm S|,
    \label{newphig}
    \end{equation}
 and to reconstruct the velocity field of each phase stored at the cell center \citep{weller2002derivation,rusche2003computational}:
    \begin{equation}
    \bm u_s = \bm u_s^* + \frac{\beta}{\mathbb D_s \rho_s}\bm u_g + \mathcal R\left [-\left(\frac{1}{\rho_s \mathbb D_s}\right)_f \nabla^{\perp} p_s |\bm S| + \left(\frac{\varepsilon_{s}}{\mathbb D_s}\right)_f \bm g\cdot \bm S  - \left(\frac{\varepsilon_s}{\rho_s \mathbb D_s}\right)_f \nabla^{\perp} p_g |\bm S|  \right ],
    \label{usnew}
    \end{equation}
    \begin{equation}
    \bm u_g = \bm u_g^* + \frac{\beta}{\mathbb D_g \rho_g}\bm u_s + \mathcal R\left [\left(\frac{\varepsilon_{g}}{\mathbb D_g}\right)_f \bm g\cdot \bm S  - \left(\frac{\varepsilon_g}{\rho_g \mathbb D_g}\right)_f \nabla^{\perp} p_g |\bm S|  \right ],
    \label{ugnew}
    \end{equation}
    where $\mathcal R$ is a linear operator, which represents the process of reconstructing the face velocity into a cell centroid value \citep{weller2014curl,weller2014non}:
    \begin{equation}
    \mathcal R(\varphi_k) = \left( \sum_f \frac{\bm {SS}}{|\bm S|} \right)^{-1} \cdot \left( \sum_f \varphi_k \frac{\bm S}{|\bm S|}  \right),
    \end{equation}
    where $\varphi_k$ is a scalar field and $\mathcal R(\varphi_k)$ becomes a vector field. This formula results from solving a minimum problem between the original flux at faces and new flux which are expressed as a function of $\bm u_k$ using a least square scheme \citep{shashkov1998local}, but employing the inverse of the face area as a weighting factor.
    When updating the velocity field at the center, the reconstruction method is used in order to eliminate the high-frequency oscillations of the pressure field.

\subsection{Solid-phase continuity equation}
A key idea of numerical solution of gas-solid flow is that although the governing equations of gas phase are solved using the gas pressure Poisson equation as usual, those of solid phase are solved using a solid volume fraction (correction) equation \citep{syamlal1998mfix,goldschmidt2001hydrodynamic,passalacqua2011ImplementationIterativeSolution}, instead of a granular pressure equation. This choice enables a much better treatment of the strong nonlinear dependency of granular pressure on solid volume fraction, and therefore a much better numerical stability \citep{van2006multiscale}. In the present study, the solid volume fraction equation can be solved either explicitly or implicitly.
An implicit solution procedure is to construct the discretized solid volume fraction equation as follows \citep{passalacqua2011ImplementationIterativeSolution}:
\begin{equation}
\frac{\partial \varepsilon_s}{\partial t} + \sum_f [(\varepsilon_s)_f \varphi_s'] - \sum_f \left[(\varepsilon_s)_f \left(\frac{1}{\mathbb D_s \rho_s}\frac{\partial p_s}{\partial \varepsilon_s}\right)_f |\bm S| \nabla^{\perp} \varepsilon_s \right] = 0,
\label{equ:continuous1}
\end{equation}
with
\begin{equation}
\varphi_s'=\varphi_s + \left( \frac{1}{\mathbb D_s \rho_s} \frac{\partial p_s}{\partial \varepsilon_s}\right)_f |\bm S| \nabla^\perp \varepsilon_s,
\label{equ:phis1'}
\end{equation}
where the expression for $\varphi_s$ depends on the selected implementation method and is summarized in Table \ref{tab:solver}. The convection term $\sum_f [(\varepsilon_s)_f \varphi_s']$ in Eq.(\ref{equ:continuous1}) is solved explicitly, and the $\sum_f \left[(\varepsilon_s)_f\left(\frac{1}{\mathbb D_s \rho_s}\frac{\partial p_s}{\partial \varepsilon_s}\right)_f |\bm S| \nabla^{\perp} \varepsilon_s \right]$ term is solved implicitly.
Note that (i) other methods for updating the solid volume fraction field are available, for example, the in-house software \textit{3D-MFM} \citep{kuipers1992numerical,van2006multiscale} and the open source software \textit{MFiX} \citep{syamlal1993mfix} solve a solid volume fraction correction equation; (ii) although we have used the word 'Implicit' here, not all are treated implicitly, for example, the term $\frac{\partial p_s}{\partial \theta_s} \nabla \theta_s$ is always treated explicitly. This is an inevitable compromise due to the use of segregated solver in present study; (iii) the treatment of $\nabla p_s = \text{snGrad}(p_s)$ in \textit{3D-MFM} and \textit{MFiX} is more implicit, as compared to the present study; and (iv) the boundedness of solid volume fraction ($0\sim\varepsilon_{s,max}$) is guaranteed by the physical nature of granular pressure.

The granular pressure gradient term can also be explicitly treated, then the equation solved becomes:
\begin{equation}
\frac{\partial \varepsilon_s}{\partial t} + \sum_f [(\varepsilon_s)_f \varphi_s] = 0.
\label{continum}
\end{equation}
In order to guarantee the boundedness of solid volume fraction, this discretized solid-phase continuity equation is rearranged as \citep{weller2002derivation}:
\begin{equation}
\frac{\partial \varepsilon_s}{\partial t} + \sum_f [(\varepsilon_s)_f \varphi] + \sum_f [(\varphi_r (\varepsilon_s)_f (\varepsilon_g)_f] = 0,
\label{rearange}
\end{equation}
where $\varphi = (\varepsilon_s)_f \varphi_s + (\varepsilon_g)_f \varphi_g$ and $\varphi_r = \varphi_s - \varphi_g$, the expressions for $\varphi_s$ are again depending on the selected implementation method and have already been summarized in Table \ref{tab:solver}.
The boundedness is ensured by employing the Multidimensional Universal Limiter with Explicit Solution (MULES) \citep{damian2014extended}, which belongs to a variant of the flux corrected transport algorithm \citep{boris1997flux}. The MULES method aims to solve the solid volume fraction equation in two steps: (i) with a first-order discretization scheme for the divergence term and (ii) use the MULES limiter and a high-order scheme to correct the divergence term. Then the solid volume flux can be limited by:
\begin{equation}
(\varepsilon_s)_f \varphi_i= (1-\lambda)[(\varepsilon_s)_f \varphi_i]^{LO} + \lambda[(\varepsilon_s)_f \varphi_i]^{HO}
\label{MULES}
\end{equation}
where $\varphi_i$ correspond to $\varphi$ and $\varphi_f$ in Eqs. (\ref{rearange}), $LO$ refers to the first-order scheme and $HO$ refers to the high-order schemes, $\lambda$ is a limiter that satisfies $0 \leq \lambda \leq 1$. More details can be found in literature \citep{zalesak1979fully,deshpande2012evaluating,phdthesisSantiago}.


It can be seen that there are implicit and explicit implementation methods and each of them has three different expressions of $\varphi_s$, therefore, there are in total six implementation methods of granular pressure gradient, and their effects on the simulation results will be studied in details and reported soon.

\setlength{\tabcolsep}{2pt}{
\small{
\begin{longtable}[H]{cc}
    \caption{Summary of the expressions for $\varphi_s$ in the solid-phase continuity equations \label{tab:solver}}
    \\\hline
    \textbf{Method for $\nabla p_s$} & \textbf{Expression of $\varphi_s$} \\
    \hline
    $\nabla p_s = \frac{\partial p_s}{\partial \varepsilon_s} \nabla \varepsilon_s$ & $\varphi_s = \left(\frac{\mathbb H_s}{\mathbb D_s}\right)_f \cdot \bm S -\left( \frac{\varepsilon_s}{\rho_s \mathbb D_s} \right)_f |\bm S| \nabla^{\perp}p_g + \left( \frac{\beta}{\rho_s \mathbb D_s} \right)_f \varphi_g + \left(\frac{\varepsilon_s}{\mathbb D_s}\right)_f \bm g \cdot \bm S - \left(\frac{1}{\rho_s \mathbb D_s}\frac{\partial p_s}{\partial \varepsilon_s} \right)_f |\bm S| \nabla^{\perp}\varepsilon_s$  \\
    $\nabla p_s = \text{snGrad}(p_s)$ & $\varphi_s = \left(\frac{\mathbb H_s}{\mathbb D_s}\right)_f \cdot \bm S -\left( \frac{\varepsilon_s}{\rho_s \mathbb D_s} \right)_f |\bm S| \nabla^{\perp}p_g + \left( \frac{\beta}{\rho_s \mathbb D_s} \right)_f \varphi_g + \left(\frac{\varepsilon_s }{\mathbb D_s} \right)_f \bm g \cdot \bm S - \left(\frac{1}{\rho_s \mathbb D_s}\right)_f \nabla^{\perp}p_s |\bm S|$ \\
    $\nabla p_s = \frac{\partial p_s}{\partial \varepsilon_s} \nabla \varepsilon_s + \frac{\partial p_s}{\partial \theta_s} \nabla \theta_s$ \ \ & $\varphi_s = \left(\frac{\mathbb H_s}{\mathbb D_s}\right)_f \cdot \bm S -\left( \frac{\varepsilon_s}{\rho_s \mathbb D_s} \right)_f |\bm S| \nabla^{\perp}p_g + \left( \frac{\beta}{\rho_s \mathbb D_s} \right)_f \varphi_g + \left(\frac{\varepsilon_s}{\mathbb D_s}\right)_f \bm g \cdot \bm S - \left(\frac{1}{\rho_s \mathbb D_s}\right)_f \left[ \left(\frac{\partial p_s}{\partial \varepsilon_s}\right)_f \nabla^{\perp} \varepsilon_s + \left(\frac{\partial p_s}{\partial \theta_s}\right)_f \nabla^{\perp} \theta_s \right] |\bm S|$ \\
    \hline
\end{longtable}
}}
\subsection{PIMPLE algorithm}
\begin{figure}
    \scriptsize
    \tikzstyle{format}=[rectangle,draw,thin]
		\tikzstyle{test}=[diamond,aspect=2,draw,thin]
		\tikzstyle{point}=[coordinate,on grid,]
		\begin{tikzpicture}[node distance =8pt]
  \node[draw, rounded corners]                        (start)   {Start};
  \node[format, below=of start]                         (initial)  {Initialization};
  \node[point,below of=initial, node distance=8pt]      (point14) {};
  \node[point,left of=point14, node distance=65mm]      (point9) {};
  \node[format, below=of initial]                         (phi_r)  {Calculate $\varphi = (\varepsilon_s)_f \varphi_s + (\varepsilon_g)_f \varphi_g$ and $\varphi_r = \varphi_s - \varphi_g$.};
  \node[point,left of=phi_r, node distance=60mm]      (point11) {};
  \node[draw, diamond, aspect=6, below =of phi_r,align=center]      (step 17) {Solve the granular pressure gradient implicitly?};
  \node[point,right of=step 17, node distance=60.4mm]      (point5) {};
  \node[format, below=15pt of step 17,align=center]                         (step 1)  {Update the granular pressure (Eq. \ref{equ::ps}) and its gradient \\(
  Method $\uppercase\expandafter{\romannumeral1}$: Eq. \ref{Ges} or \ref{Ges2}, Method $\uppercase\expandafter{\romannumeral3}$: Eqs. \ref{Ges} or \ref{Ges2} and \ref{equ::pstheta}).};
  \node[format, below=of step 1,align=center]                         (phi_r')  {Calculate volumetric fluxes $\varphi'$ and $\varphi_r'$ without contribution of granular pressure gradient,\\$\varphi'=\varphi' + (\varepsilon_s)_f\left( \frac{1}{\mathbb D_s \rho_s} \frac{\partial p_s}{\partial \varepsilon_s}\right)_f |\bm S| \nabla^\perp \varepsilon_s$, $\varphi_r'=\varphi_r'+\varphi_s'=\varphi_s + \left( \frac{1}{\mathbb D_s \rho_s} \frac{\partial p_s}{\partial \varepsilon_s}\right)_f |\bm S| \nabla^\perp \varepsilon_s$.};
  \node[format, below =of phi_r',align=center]                        (step 25) {Solve the solid-phase continuity equation using\\the MULES limiter $\frac{\partial \varepsilon_s}{\partial t} + \sum_f[ (\varepsilon_s)_f \varphi_s' ] = 0$,\\where $\varphi'_s$ is calculated based on $\varphi'$ and $\varphi_r'$. (Eqs. \ref{equ:phis1'}, \ref{MULES}, Table. \ref{tab:solver}).};
  \node[point,left of=step 25, node distance=40mm]      (point20) {};
  \node[draw, diamond, aspect=3, below =of step 25,align=center]      (step 26) {Pre-determined number of\\ correction is reached?};
  \node[point,left of=step 26, node distance=40mm]      (point21) {};
  \node[format, right=of step 25,align=center]                        (step 2)  {Solve the solid-phase continuity equation using\\the MULES limiter,where\\$\varphi_s$ is calculated based on $\varphi$ and $\varphi_r$\\(Eqs. \ref{rearange}, \ref{MULES}, Table. \ref{tab:solver}).};
  \node[point,right of=step 2, node distance=30mm]      (point13) {};

  \node[draw, diamond, aspect=3, below =of step 2,align=center]      (step 4) {Pre-determined number of\\ correction is reached?};
  \node[point,right of=step 4, node distance=30mm]      (point2) {};
  \node[format, below =15pt of step 26]                        (step 18) {Solve the solid-phase continuity equation using an implicit method (Eqs. \ref{equ:continuous1}, \ref{equ:phis1'}, Table. \ref{tab:solver}).};
  \node[point,below of=step 18, node distance=9pt]      (point6) {};
  \node[point,right of=point6, node distance=60.4mm]      (point22) {};
  \node[format, below =of step 18]                        (step 3) {Obtain the continuous phase fraction as $1-\varepsilon_s$.};
  \node[format, below= of step 3]                        (step 5) {Update the drag coefficients with the new value of the phase fractions (Eqs. \ref{equ::EMMSbeta}-\ref{Gidaspowbeta}).};

  \node[format, below=of step 5]                        (step 7) {Compute phase stress tensors (Eqs. \ref{taug},\ref{taus}).};
  \node[format, below=of step 7,align=center]                        (step 9) {Estimate the phase pseudo-velocities of gas and solid phases\\according to the assembled phase momentum equation (Eq. \ref{u*}).};
  \node[format, below=of step 9,align=center]                        (step 11) {Compute the new phase volumetric fluxes $\varphi^0$, $\varphi^0_s$, $\varphi^0_g$, which do not\\include the contribution of the gas pressure gradient (Eqs. \ref{phi0}-\ref{phif0}).};
  \node[point,left of=step 11, node distance=55mm]      (point12) {};
  \node[format, below=of step 11]                        (step 12) {Construct and solve the gas pressure equation (Eq. \ref{Poisson}).};
  \node[point,left of=step 12, node distance=50mm]      (point8) {};
  \node[format, below=of step 12]                        (step 14) {Correct the phase volumetric fluxes (Eqs. \ref{newphis},\ref{newphig}).};
  \node[format, below=of step 14]                        (step 15) {Correct the phase velocities using the flux reconstruction procedure (Eqs. \ref{usnew},\ref{ugnew}).};
  \node[draw, diamond, aspect=6, below =of step 15,align=center]      (step 13) {Pre-determined number of correction is achieved?};
  \node[point,left of=step 13, node distance=50mm]      (point3) {};
  \node[draw, diamond, aspect=6, below =15pt of step 13,align=center]                        (step 23) {Pre-defined number of correction is achieved?};
  \node[point,left of=step 23, node distance=55mm]      (point10) {};
  \node[draw, diamond, aspect=4, below =15pt of step 23,align=center]                        (step 24) {Pre-defined number of correction is achieved?};
  \node[point,left of=step 24, node distance=60mm]      (point15) {};
  \node[format, below=15pt of step 24]                        (step 6) {Solve the granular temperature equation (Eq. \ref{equ::granulartemperature}).};
  \node[draw, diamond, aspect=4, below = of step 6,align=center]                        (step 21) {Pre-defined number of time step is achieved? };
  \node[point,left of=step 21, node distance=65mm]      (point7) {};
  \node[draw, rounded corners, below=15pt of step 21]         (end)     {End};
  \draw[-latex] (start)  -- (initial);
  \draw[-latex] (initial)  -- (phi_r);
  \draw[-latex] (phi_r)  -- (step 17);
  \draw[-latex] (step 17) -- node[left]  {Yes} (step 1);
  \draw[-] (step 17) -- node[above] {No} (point5);
  \draw[-latex] (point5) -- (step 2);
  \draw[-latex] (step 1) -- (phi_r');
  \draw[-latex] (phi_r') -- (step 25);
  \draw[-latex] (step 25)  -- (step 26);
  \draw[-latex] (step 26) -- node[left]  {Yes} (step 18);
  \draw[-] (step 26) -- node[above] {No} (point21);
  \draw[-] (point21) -- node[left]{\rotatebox{90}{MULES loop}} (point20);
  \draw[-latex](point20) -- (step 25);

  \draw[-latex] (step 2) -- (step 4);
  \draw[-] (step 4) -- node[right]  {Yes} (point22);
  \draw[-latex] (point22) --  (point6);
  \draw[-] (step 4) -- node[above] {No} (point2);
  \draw[-] (point2) -- node[right]{\rotatebox{270}{MULES loop}} (point13);
  \draw[-latex](point13) -- (step 2);
  \draw[-latex] (step 18) -- (step 3);
  \draw[-latex] (step 3) -- (step 5);
  \draw[-latex] (step 5) -- (step 7);

  \draw[-latex] (step 7) -- (step 9);
  \draw[-latex] (step 9) -- (step 11);
  \draw[-latex] (step 11) -- (step 12);
  \draw[-latex] (step 12) -- (step 14);
  \draw[-latex] (step 13) -- node[left]  {Yes} (step 23);
  \draw[-] (step 13) -- node[above] {No} (point3);
  \draw[-] (point3) -- node[left]{\rotatebox{90}{Non-orthogonal correction}} (point8);
  \draw[-latex](point8) -- (step 12);
  \draw[-latex] (step 14) -- (step 15);
  \draw[-latex] (step 15) -- (step 13);
  \draw[-latex] (step 23) -- node[left]  {Yes} (step 24);
  \draw[-] (step 23) -- node[above] {No} (point10);
  \draw[-] (point10) -- node[left]{\rotatebox{90}{PIMPLE correction loop}} (point12);%
  \draw[-latex](point12) -- (step 11);
  \draw[-latex] (step 24) -- node[left]  {Yes} (step 6);
  \draw[-] (step 24) -- node[above] {No} (point15);
  \draw[-] (point15) -- node[left]{\rotatebox{90}{PIMPLE outer-correction loop}} (point11);
  \draw[-latex](point11) -- (phi_r);
  \draw[-latex] (step 6) -- (step 21);
  \draw[-latex] (step 21) -- node[left]  {Yes} (end);
  \draw[-] (step 21) -- node[above] {No} (point7);
  \draw[-] (point7) -- node[left]{\rotatebox{90}{Time loop}} (point9);%
  \draw[-latex](point9) -- (point14);
    \end{tikzpicture}
    \caption{Velocity-pressure coupling algorithm (PIMPLE algorithm).}
    \label{PIMPLE}
	\end{figure}
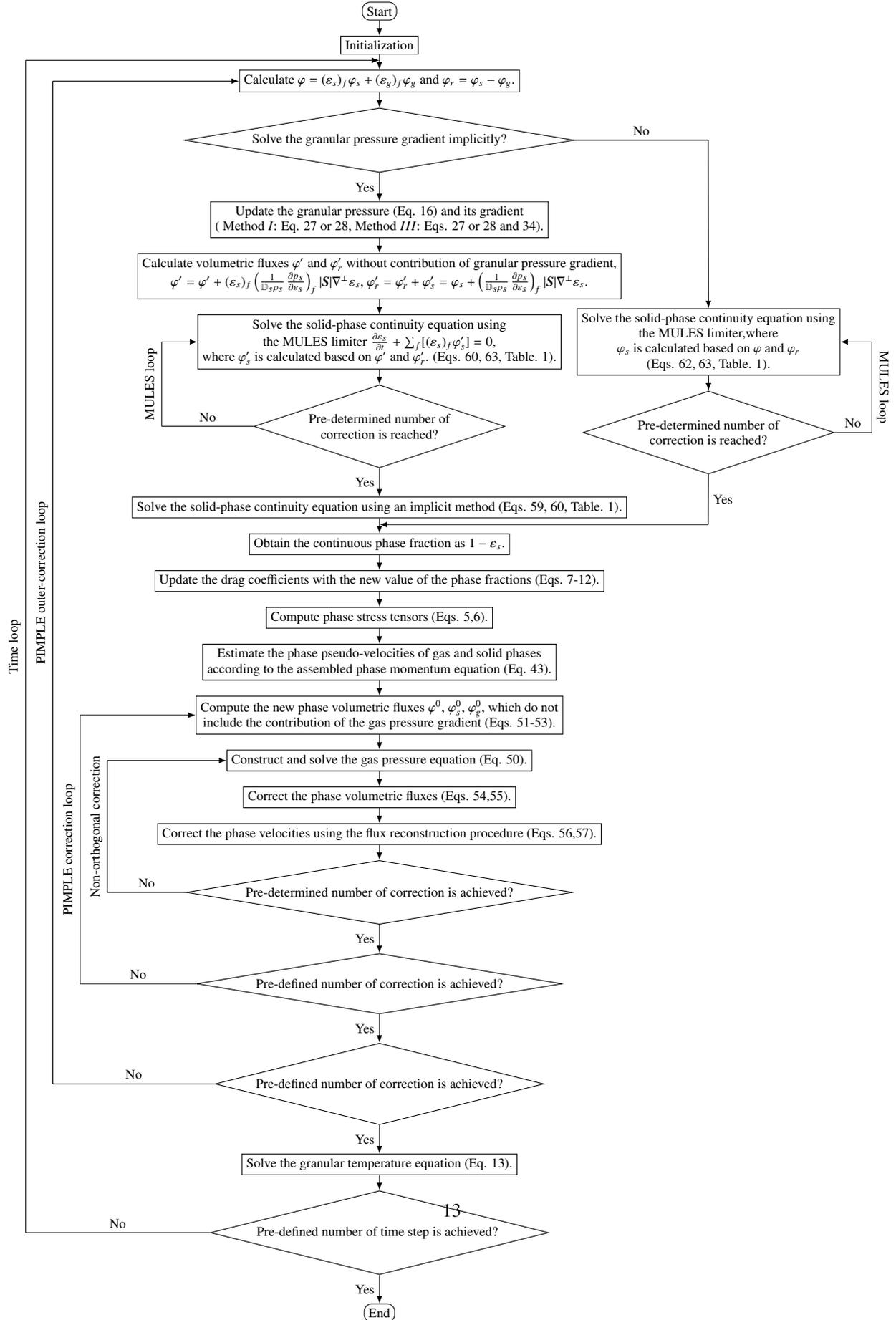

The solution of governing equations for Euler-Euler two-fluid model is complex due to the strong coupling between unknown variables,  additionally, the gas pressure appears in both of the gas and solid momentum conservation equations, but no transport equation for the gas pressure is obviously available, and finally, the granular pressure is a strong nonlinear function of solid volume fraction. To solve those issues, the equations are transformed into phasic momentum equations, gas pressure equation and solid-phase continuum equation as have been described in details in previous sections, and then tackled through a segregated or an iterative algorithm. The segregated algorithm is a combination of PISO (Pressure Implicit with Splitting of Operators) \citep{issa1986solution} and SIMPLE (Semi-Implicit Method for Pressure-Linked Equations) \citep{patankar1972Calculation}, which is called PIMPLE (Pressure-Implicit Method for Pressure-Linked Equations) in OpenFOAM$^{\circledR}$  \citep{weller2002derivation}. This PIMPLE algorithm is summarized in Fig. \ref{PIMPLE} and explained in the following paragraphs.

The initial and boundary conditions for all field values should firstly be specified, the calculation of the new time-step value can then be started.
Next, the solid phase continuity equation should be solved. It is worth noting that: (i) Eq. (\ref{rearange}) is a nonlinear function of $\varepsilon_s$. Therefore, boundedness at both ends can be guaranteed only if the equation is solved fully implicitly, with a proper discretization scheme for the convective term. Thus, solution with linear solvers requires iteration, which is done in the \textbf{MULES correction loop}. Another way to address this issue is by utilizing a small Courant number, which is easily achievable for transient problems since it inherently involves small time steps. (ii) To avoid surpassing the packing limit of solid volume fraction. One feasible approach is to deal with the granular pressure implicitly in the solid volume fraction equation. Another effective solution is to employ the MULES correction, which not only secures this objective but also prevents any dramatic change of solid volume fraction occurring. Note that the form of the solid phase continuity equation solved by MULES approach varies with implicit versus explicit methods.

The momentum transfer coefficients and phase stress tensor are then updated using the new phase volume fraction, this process is followed by solving the granular temperature equation. The incomplete phase momentum equation can then be assembled, with the next step being the \textbf{momentum predictor}. In this step, the phase pseudo-velocities are initially estimated by applying the assembled incomplete phase momentum equations through Eq. (\ref{u*}), where $\mathbb H_k$ is evaluated by re-substituting the previously obtained velocities into their discretized equations, to ensure a solution without any needs of iterations. The $\bm u_k^*$ term composes of two parts \citep{jasak1996error}: the "transport part", which incorporates the matrix coefficients for all neighbouring cells multiplied by corresponding the phase velocities and the "source part" that arises from the transient term.
It has been discovered that the solution of the momentum predictor may destabilize the solution procedure in challenging circumstances. As a result, the discretized incomplete momentum equations are only constructed but not solved, which restricts the time step in transient calculations. However, so far, this has not been identified as a significant limitation \citep{rusche2003computational}. On the other hand, it should also be noted that not solving the momentum predictor can impede the convergence of steady-state calculations.

Next is the \textbf{PIMPLE correction loop}. In this step, additional source terms in the phase momentum equations, namely, the gravity term, explicit part of the drag term and granular pressure term are included apart from that the gas pressure term is supplemented in Eqs. (\ref{phi0}-\ref{phif0}). Rhie-Chow interpolation \citep{rhie1983numerical} is adopted to construct the predicted phase volumetric fluxes at this stage.
By combining the predicted phase volumetric fluxes with the phase continuity equations, the gas pressure equation can be formulated. The solution of gas pressure equation provides an initial estimate of the new pressure field.
\begin{figure}[!htb]
    \centering
    \includegraphics[scale=1]{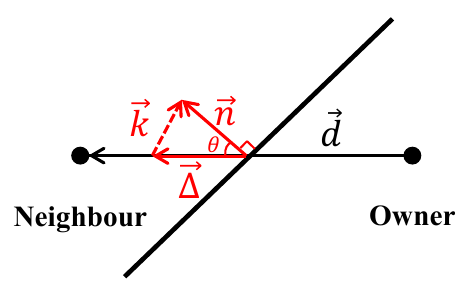}
    \caption{Decomposition schematic of the face normal vector.}
    \label{nonorthogonal}
\end{figure}
There is a critical problem when solving the gas pressure equation. When the geometric meshes are not completely orthogonal, as shown in Fig. \ref{nonorthogonal}, where $\vec d$ is the vector pointing from the center of Owner cell to the center of Neighbour cell. The surface normal gradient of the pressure $\nabla^{\perp}p_g$ in the direction of $\vec n$ can be decomposed into the orthogonal component $\vec \Delta$ parallel to $\vec d$ and the non-orthogonal component $\vec k$.
\begin{equation}
\vec n = \vec \Delta + \vec k.
\end{equation}
The value of the orthogonal component $\vec \Delta$ depends on the decomposition method. In this study, the over-relaxed approach is used \citep{greenshieldsweller2022}, since it is able to decompose the non-orthogonal part to a minimum compared to other approaches and works best \citep{jasak1996error}:
\begin{equation}
\vec \Delta = \frac{|\vec n|}{cos \theta} \frac{\vec d}{|\vec d|},
\end{equation}
where $\theta$ is the angel between $\vec \Delta$ and $\vec n$. Then the gradient of gas pressure at the face becomes:
\begin{equation}
\nabla^{\perp} p_g = \vec \Delta \frac{p_{g,P}-p_{g,N}}{|\bm d_P - \bm d_N|}+\vec k (\nabla p_g)_f
\label{equ:nonorthogonal}
\end{equation}
where the surface interpolation of $(p_g)_f$ is calculated as same as Eq. (\ref{interpolate}). The implicit computation of the orthogonal term $\vec \Delta \frac{p_{g,P}-p_{g,N}}{|\bm d_P - \bm d_N|}$ in Eq. (\ref{equ:nonorthogonal}) includes only the first neighbors of the cell and results in a diagonally equal matrix. In contrast, if the non-orthogonal correction term $\vec k (\nabla p_g)_f$ in Eq. (\ref{equ:nonorthogonal}) is implicitly approximated, the "second neighbors" of the control volume into the computational molecule with negative coefficients would be introduced, violating diagonal equality, and possibly causing unboundedness, especially when the mesh non-orthogonality is high. If ensuring boundedness is of priority over accuracy, it may be necessary to limit or completely eliminate the non-orthogonal correction term \citep{rusche2003computational}. Therefore, the non-orthogonal correction is usually explicitly treated and added into the corresponding source vector \citep{greenshieldsweller2022}. Consequently, the gas pressure equation (Eq. (\ref{Poisson})) must be solved multiple times, with each solution updating the non-orthogonal correction term, until the desired tolerance is met or the pre-defined number of iterations is reached. This approach will enable the acquisition of a cell-centered new pressure field. This is the \textbf{non-orthogonal correction} of the gas pressure equation. In short, implementing this practice will enhance the quality of the coefficient matrix. However, non-orthogonal correction is typically not necessary unless the non-orthogonal angle exceeds 70 degree. Even in such cases, the number of correction should not surpass two.

The volumetric fluxes for each phase that align with the new pressure field can be updated using Eqs. (\ref{newphis},\ref{newphig}). Additionally, the phase velocity fields should be adjusted to accommodate the new pressure distribution. These velocity adjustments are made explicitly using Eqs. (\ref{usnew},\ref{ugnew}), representing the explicit velocity correction stage. The reconstruction method is employed to eliminate high-frequency oscillations of the updated fields. Upon further examination of Eqs. (\ref{usnew},\ref{ugnew}), it becomes apparent that the correction to the phase velocity is comprised of multiple components. One such component is the correction resulting from the alteration of the gas pressure gradient term $\mathcal R[-(\frac{\varepsilon_k}{\rho_k \mathbb D_k})_f \nabla^{\perp} p_g |\bm S|]$ in addition to other terms. The explicit phase velocity correction implies the omission of all terms aside from the gas pressure term. Consequently, assuming that the velocity error originates exclusively from an error in the gas pressure term is unreasonable. Correction of other terms, development of a new gas pressure equation, and repeating the process is therefore required. In essence, the \textbf{PIMPLE correction loop} comprises a series of pressure solution and explicit phase velocity corrections, which are repeated until meeting a pre-determined tolerance or iteration number, typically 2-3.

Another concern arises from the fact that the $\mathbb H_k$ coefficients depended on the phase flux fields. Following each solution, a brand new set of conservative fluxes becomes available, offering a chance to recompute the coefficients in $\mathbb H_k$. Nevertheless, this step is disregarded, as it is believed that the nonlinear coupling is less significant than the pressure-velocity coupling, which is in line with the linearization of the momentum equation. The coefficients in $\mathbb H_k$ remain the same throughout the correction sequence and will not be altered until the next \textbf{momentum predictor}. Fortunately, the inconsistency among the phase volume fraction, coefficient matrix, phase velocities and gas pressure can be resolved by setting a reasonable number of \textbf{PIMPLE outer-correction loop}.
Once the \textbf{PIMPLE outer-correction loop} is completed, the phase volume fraction, the gas pressure and phase velocity fields for the current time-step are obtained, along with the new set of phase conservative fluxes.

Then a new cycle of iterations commences and is repeated until a pre-defined time step is reached. It is worth noting that adopting the under-relaxation method can enhance the convergence and robustness of the numerical solution.

\clearpage
\section{Test cases}
Present article focuses on studying the effects of the six different implementation methods of granular pressure gradient on the simulation results of Euler-Euler model, as summarized in Section 3, each of which can be solved either explicitly or implicitly. Therefore, the model summarized in Section 2 is numerically solved, using the algorithm reported in Section 4, to simulate the hydrodynamics of gas-solid flows in two bubbling fluidized beds and one circulating fluidized bed riser.

    \subsection{Bubbling fluidized beds}
        \subsubsection{Simulation setup}
A two-dimensional bubbling fluidized bed was simulated as case $\uppercase\expandafter{\romannumeral1}$, which was experimentally studied by \cite{taghipour2005experimental}, and also numerically studied by \citet{taghipour2005experimental} and \citet{herzog2012comparative}.
Another two-dimensional bubbling fluidized bed with Geldart B particles, which has been numerically examined by \citet{parmentier2008numerical}, \citet{passalacqua2011ImplementationIterativeSolution} and \citet{venier2016numerical}, was simulated as case $\uppercase\expandafter{\romannumeral2}$.
Tables. \ref{tab:conf2} and \ref{tab:conf3} summarize the parameters used in the simulations respectively. We have set the parameters as possible as same to the corresponding numerical solutions \citep{taghipour2005experimental,herzog2012comparative, passalacqua2011ImplementationIterativeSolution,parmentier2008numerical,venier2016numerical}, although differences may still exist since some parameters were not reported in previous studies.
The boundary conditions are presented in Table. \ref{tab:boundary2}, the numerical schemes are presented in Table. \ref{tab:scheme2}, and the under-relaxation factors are presented in Table. \ref{tab:relax2}. The simulations are carried out for 30 s real time and data are time-averaged for the last 25 s for both bubbling fluidized beds $\uppercase\expandafter{\romannumeral1}$ and $\uppercase\expandafter{\romannumeral2}$ for postprocessing. For case $\uppercase\expandafter{\romannumeral1}$, the time step is adapted according to the Courant number which is defined as $Co=\frac{\Delta t |\bm u_r|}{\Delta x}$,
where $\Delta t$ is the time step and $|\bm u_r|$ is the magnitude of the interphase slip velocity. The Courant number is not permitted to be larger than 0.1, with the pre-defined maximum of time step set to $1\times 10^{-4} s$.
For the bubbling fluidized bed $\uppercase\expandafter{\romannumeral2}$, the fixed time step is $1\times 10^{-4} s$.

\begin{longtable}[H]{lc}
    \caption{Summary of parameters used in the simulation of bubbling fluidized bed $\uppercase\expandafter{\romannumeral1}$ \label{tab:conf2}}
    \\\hline
    \textbf{Properties} & \textbf{Value} \\
    \hline
    Bubble fluidized bed diameter $D(m)$                   & 0.28 \\
    Bubble fluidized bed height $H(m)$                     & 1.0 \\
    Particle diameter $d_p(\mu m)$              & 275 \\
    Particle density $\rho_s(kg/m^3)$       & 2500 \\
    Gas density $\rho_g(kg/m^3)$            & 1.225 \\
    Gas viscosity $\mu_g(kg/m\cdot s)$      & $1.485\times10^{-5}$ \\
    Superficial gas velocity $U_g(m/s)$     & 0.03 \quad 0.1 \quad 0.13 \quad 0.17 \quad 0.2 \quad 0.38 \quad 0.46 \quad 0.51 \quad 0.6 \\
    Initial solids packing $\varepsilon_{s,ini}$      & 0.6 \\
    Static bed height $H_0(m)$              & 0.4   \\
    Particle-particle restitution coefficient & 0.9\\
    Particle-wall restitution coefficient & 0.9\\
    Specularity coefficient & 0.5\\
    Grid size (mm$\times$mm)& 5$\times$5\\
    \hline
\end{longtable}
\begin{longtable}[H]{lc}
    \caption{Summary of parameters used in the simulation of bubbling fluidized bed $\uppercase\expandafter{\romannumeral2}$ \label{tab:conf3}}
    \\\hline
    \textbf{Properties} & \textbf{Value} \\
    \hline
    Bubble fluidized bed diameter $D(m)$                   & 0.138 \\
    Bubble fluidized bed height $H(m)$                     & 1.0 \\
    Particle diameter $d_p(\mu m)$              & 350 \\
    Particle density $\rho_s(kg/m^3)$       & 2000 \\
    Gas density $\rho_g(kg/m^3)$            & 1.4 \\
    Gas viscosity $\mu_g(kg/m\cdot s)$      & $1.8\times10^{-5}$ \\
    Superficial gas velocity $U_g(m/s)$     & 0.54 \\
    Average voidage in bed $\varepsilon_{s,bed}$      & 0.116 \\
    Particle-particle restitution coefficient & 0.8\\
    Particle-wall restitution coefficient & 0.8\\
    Specularity coefficient & 0.1\\
    Grid number (mm$\times$mm)              & $1\times1$\\
    \hline
\end{longtable}

\begin{longtable}[H]{cccc}
    \caption{Summary of boundary conditions used in the simulations of bubbling fluidized bed $\uppercase\expandafter{\romannumeral1}$ and $\uppercase\expandafter{\romannumeral2}$. \label{tab:boundary2}}
    \\\hline
    \textbf{Variables} & \textbf{gas\_inlet} & \textbf{outlet} & \textbf{walls}\\
    \hline
    $\alpha_s$ & zeroGradient & zeroGradient & zeroGradient  \\
    $\bf {U_s}$ & 0 & 0 & JohnsonJacksonParticleSlip \\
    $\bf {U_g}$ & fixedValue & pressureInletOutletVelocity &  noslip \\
    $p$ & zeroGradient & 101325 Pa & zeroGradient \\
    $\theta_s$ & 0 & zeroGradient & JohnsonJacksonParticleTheta \\
    \hline
\end{longtable}

\begin{longtable}[H]{lcccc}
    \caption{Summary of numerical schemes used in the simulations of bubbling fluidized bed $\uppercase\expandafter{\romannumeral1}$ and $\uppercase\expandafter{\romannumeral2}$. \label{tab:scheme2}}
    \\\hline
    \textbf{Term} & \textbf{Scheme} \\
    \hline
    $\partial / \partial t$ & bubbling fluidized bed $\uppercase\expandafter{\romannumeral1}$: Euler \\
    \quad & bubbling fluidized bed $\uppercase\expandafter{\romannumeral2}$: backward \\
    $\nabla \psi$ & Gauss linear \\
    $\nabla \cdot \psi$ & Gauss limitedLinear \\
    $\nabla \cdot (\nabla \psi)$ & Gauss linear corrected \\
    $\nabla^{\perp}$ & Corrected \\
    $(\psi)_f$ & Gauss linear \\
    \hline
\end{longtable}

\begin{longtable}[H]{lc}
    \caption{Summary of under-relaxation factors used in the simulations of bubbling fluidized bed $\uppercase\expandafter{\romannumeral1}$ and $\uppercase\expandafter{\romannumeral2}$. \label{tab:relax2}}
    \\\hline
    \textbf{Variables} & \textbf{Under-relaxation Value} \\
    \hline
    $p$ & 0.3  \\
    $\bm u_k$      & 0.7  \\
    $\varepsilon_k$      & 0.2 \\
    $\theta_s$   & 0.2 \\
    \hline
\end{longtable}

\subsubsection{Bubbling fluidized bed $\uppercase\expandafter{\romannumeral1}$}

\begin{figure}[!htb]
    \centering
    \includegraphics[scale=0.5]{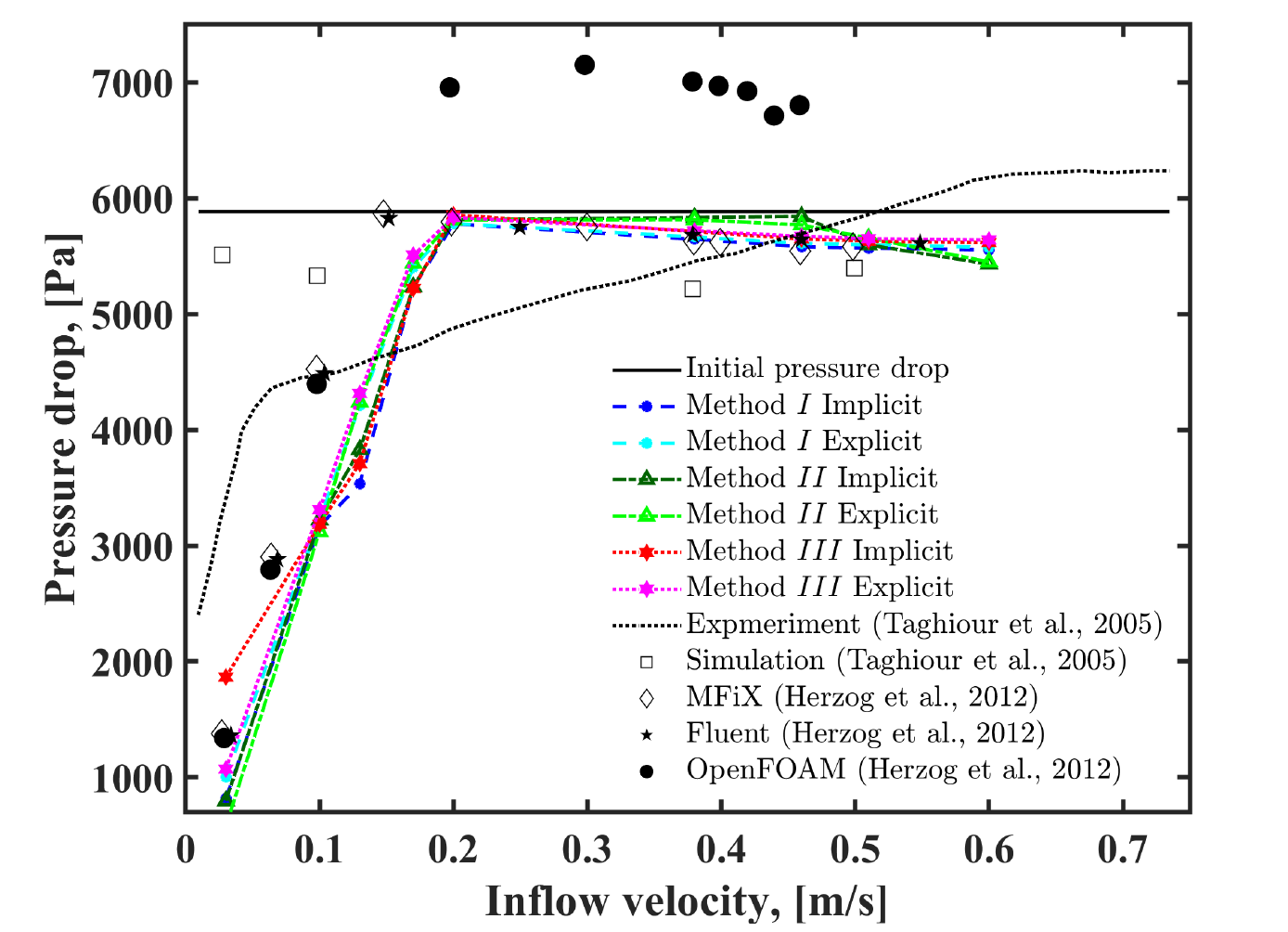}
    \caption{Time-averaged pressure drop values as a function of inflow velocity in the bubbling fluidized bed $\uppercase\expandafter{\romannumeral1}$, and comparison with the simulation results of \cite{taghipour2005experimental} and \cite{herzog2012comparative}, and the experimental data of \cite{taghipour2005experimental}.}
    \label{pU}
\end{figure}

Numerical analysis were conducted to examine the steady state pressure drop $\Delta p$ and the bed expansion ratio $H/H_0$ under various superficial gas velocities. For the simulations in this study, different superficial gas velocities from $0.03 m/s$ up to $0.6 m/s$ were investigated. Fig. \ref{pU} compares the time-averaged pressure drop values against the inflow velocities (or the superficial gas velocities) with the existing measurements data \citep{taghipour2005experimental} as well as numerical solutions \citep{herzog2012comparative,taghipour2005experimental}.
First of all, all simulations can not well reproduce the experimentally measured pressure drops of \citet{taghipour2005experimental}, the underlying reasons remain unclear. It however needs to be pointed out that the experimentally measured pressure drop as a function of superficial gas velocity did not follow a common sense of gas fluidization of Geldart B particles. Moreover, \citet{taghipour2005experimental} determined that the minimum fluidization velocity to be $\bm U_{mf}=0.065m/s$ from the reported pressure drops.
When $\bm U_g < \bm U _{mf}$, the Implicit outcome yielded by method $\uppercase\expandafter{\romannumeral3}$ differ from other methods. As $\bm U_g$ increases and surpasses $\bm U_{mf}$, all methods in this study provide consistent results. \citet{taghipour2005experimental} observed that the Syamlal-O'Brien drag model would overestimate the minimum fluidization velocity by a factor of three, specifically, $\bm U_{mf}$=0.20m/s, which matches the predicted fluidization velocity in this study. Failure of predicting the accurate minimum fluidization velocity can be remedied by either utilizing the Gidaspow drag model instead, as has been investigated by \citet{herzog2012comparative}, or modifying the Syamlal-O'Brien drag model by changing its constants from 0.8 and 2.65 to 0.28 and 9 \citep{syamlal2003fluid}. Since the primary focus of this study was to examine the effects of different implementation methods of the granular pressure gradient term, and the comparison to previous simulations using identical numerical models, these endeavors were not carried out in present study.
The pressure drop derived from OpenFOAM$^{\circledR}$ by \citet{herzog2012comparative} eventually stabilizes at another higher plateau that is significantly different with the initial pressure drop or the particle weight ($\Delta p = \overline{\varepsilon_{s,ini}}\rho_s \bm g$). Since it is well accepted that the pressure drop should be approximately equal to the particle weight when they are fully fluidized, this observation is a clear indication that the version of OpenFOAM$^{\circledR}$ used by \citet{herzog2012comparative} has certain major defects, or else, the initial conditions have not been given properly.

\begin{figure}[!htb]
    \centering
    \includegraphics[scale=0.5]{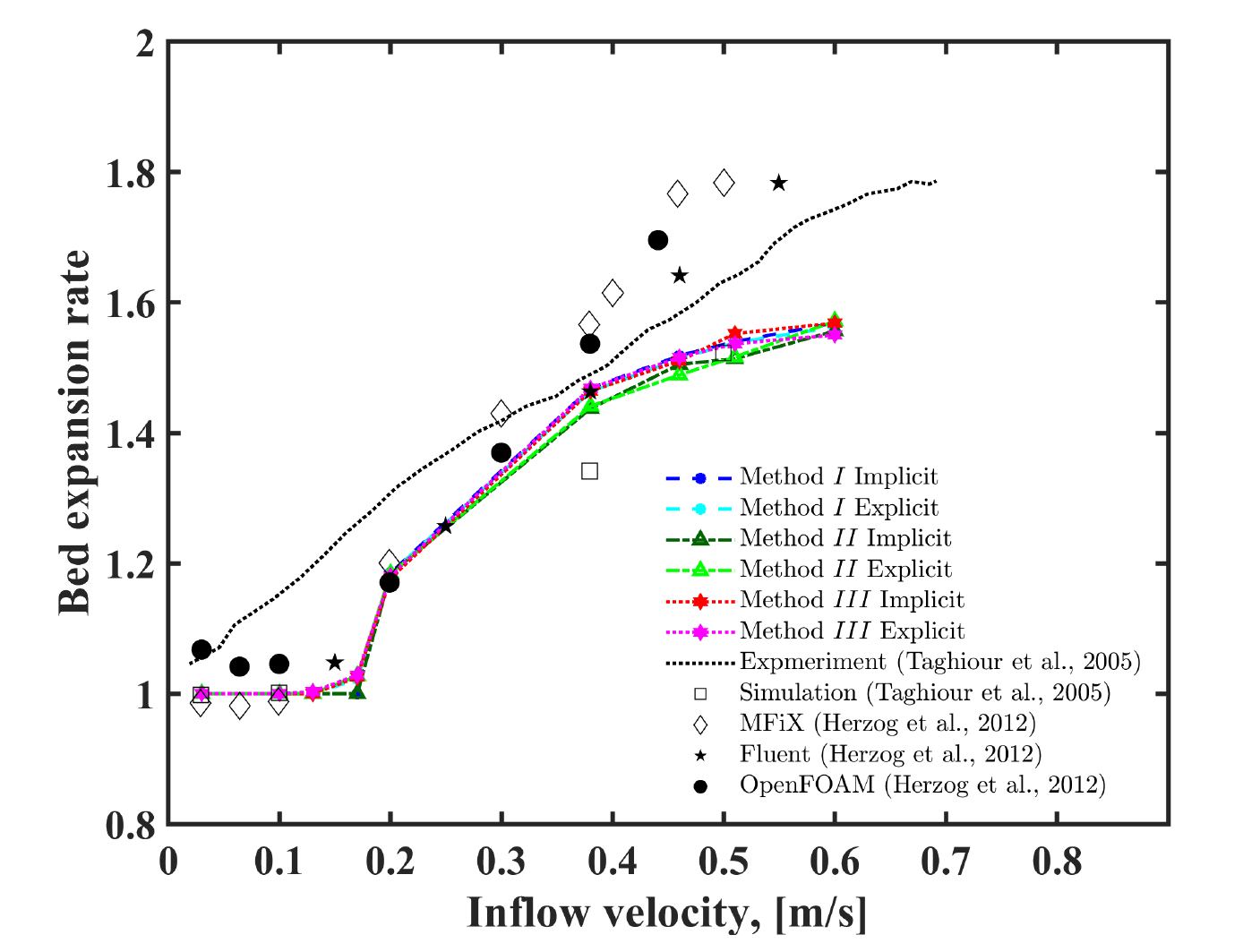}
    \caption{Time-averaged bed expansion rate as a function of inflow velocity in the bubbling fluidized bed $\uppercase\expandafter{\romannumeral1}$, and comparison with the simulation results of \cite{taghipour2005experimental} and \cite{herzog2012comparative}, and the experimental data of \cite{taghipour2005experimental}.}
    \label{HU}
\end{figure}

Fig. \ref{HU} presents the time-averaged bed expansion rate against inflow gas velocity, along with other numerical and experimental data. The bed expansion rate was computed using the pressure drop along a vertical midline of the bed, which is equivalent to \citet{herzog2012comparative}. All six methods consistently demonstrate that bed expansion increases with inflow gas velocity. Additionally, our simulations predict bed height accurately at lower gas velocities but underestimate bed expansion rate at higher gas velocities compared to simulations \citet{herzog2012comparative} and experiment \citep{taghipour2005experimental}. The disparities for $\bm U_g < \bm U_{mf}$ compared to \citet{herzog2012comparative} conducted on OpenFOAM$^{\circledR}$ may due to the different frictional forces within particles, which remains unclear in previous simulations \citep{taghipour2005experimental,herzog2012comparative}.

\begin{figure}[!htb]
    \centering
    \includegraphics[scale=0.48]{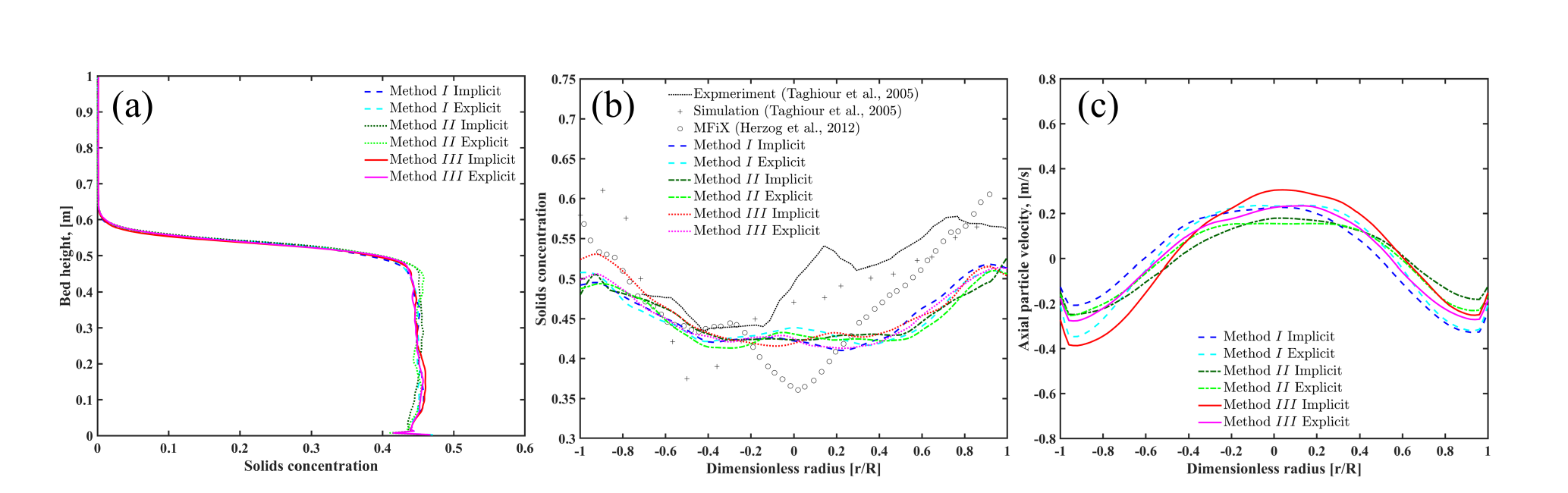}
    \caption{Time-averaged profiles for $\bm U_g = 0.38m/s$ with (a): axial solids concentration, (b): radial solids concentration at the height of 0.2m, (c): radial particle velocity at the height of 0.2 m, with comparison to the simulation results \citep{herzog2012comparative,taghipour2005experimental}, and experimental data \citep{taghipour2005experimental}.}
    \label{Tagh038}
\end{figure}
\begin{figure}[!htb]
    \centering
    \includegraphics[scale=0.48]{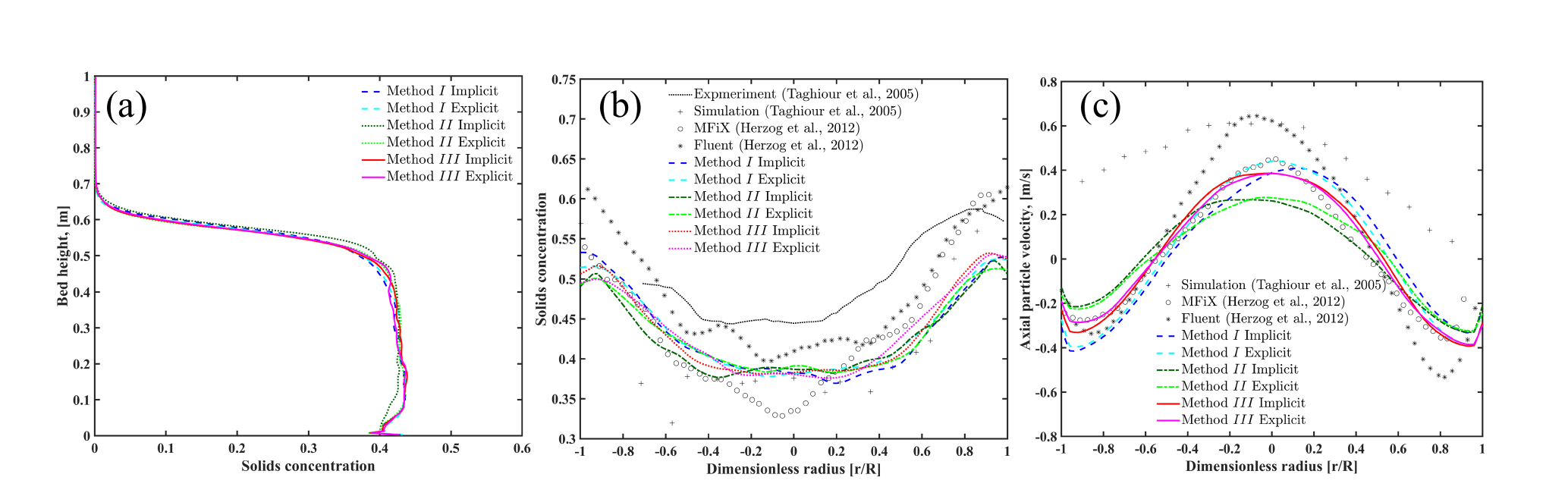}
    \caption{Time-averaged profiles for $\bm U_g = 0.46m/s$ with (a): axial solids concentration, (b): radial solids concentration at the height of 0.2m, (c): radial particle velocity at the height of 0.2 m, with comparison to the simulation results \citep{herzog2012comparative,taghipour2005experimental}, and experimental data \citep{taghipour2005experimental}.}
    \label{Tagh046}
\end{figure}

Figs. \ref{Tagh038}, \ref{Tagh046} report the time-averaged profiles for $\bm U_g$=0.46m/s and $\bm U_g$=0.38m/s, respectively. The results are identical for all six methods tested across the two gas velocities.
The radial solids concentration and velocity profiles of this study demonstrate superior symmetry to the published numerical solutions and measurements, likely due to the steady state data being collected over a longer period of time, ranging from 5-30 s. In contrast, the asymmetry of the results from \citet{taghipour2005experimental} can be attributed to the limited time frame for data collection. Additionally, the data from \citet{herzog2012comparative} was collected for 3-12 s, which is too short to reach symmetry, despite reaching stabilization of the overall bed pressure drop after about 3 s \citep{taghipour2005experimental}.
According to the radial particle velocity distributions, our simulations provide upward particle velocity within the bed interior and a downward flow in the vicinity of the side walls, as also predicted by \citet{herzog2012comparative}. This is a typical particle velocity profile due to the extensive bubbling structure. Profiles at $\bm U_g$=0.46m/s also exhibit great accordance with \citet{herzog2012comparative} using MFiX.

    \subsubsection{Bubbling fluidized bed $\uppercase\expandafter{\romannumeral2}$}
\begin{figure}[!htb]
    \centering
    \includegraphics[scale=0.5]{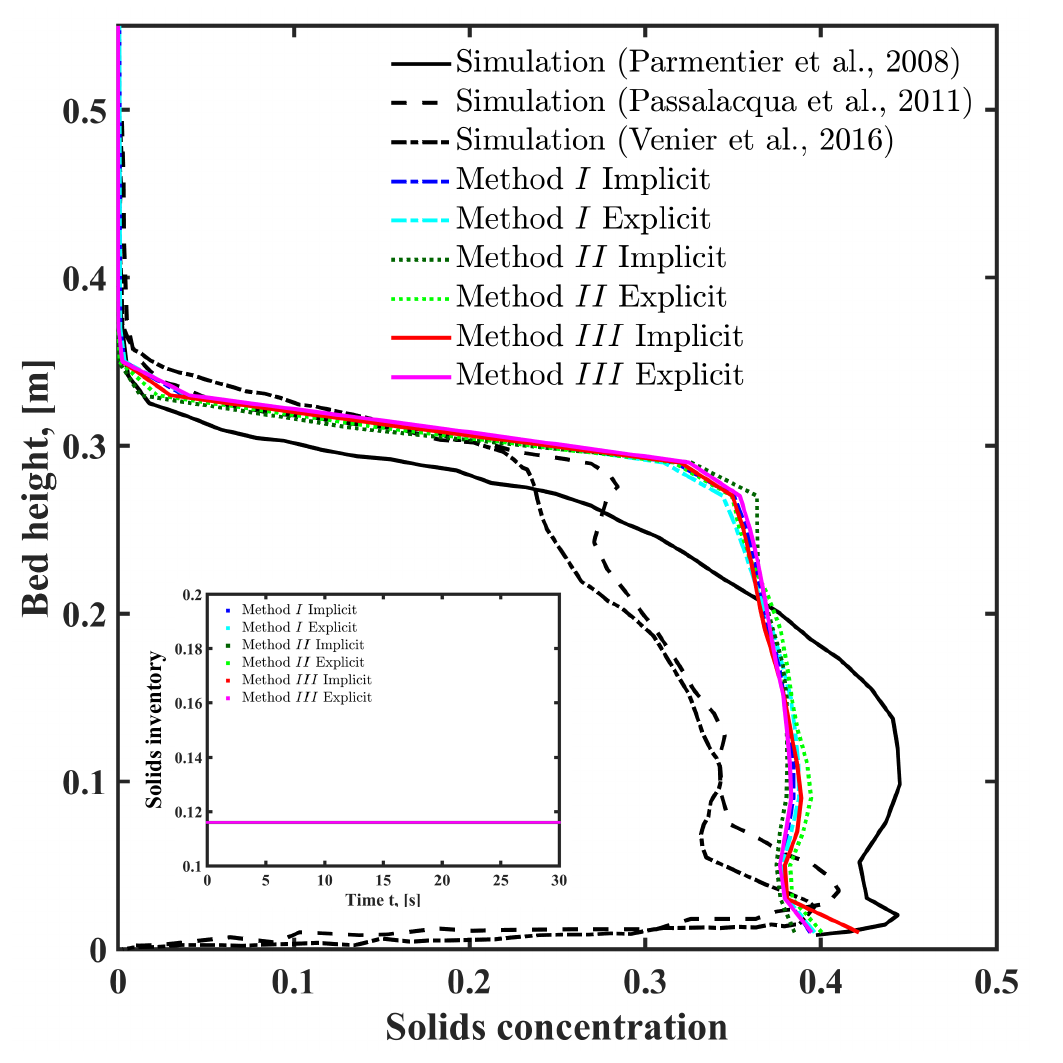}
    \caption{Time-averaged axial profile of solids concentration, with a comparison to the simulation results of \citet{passalacqua2011ImplementationIterativeSolution}, \citet{parmentier2008numerical} and \citet{venier2016numerical}, with an accompanying demonstration of the time evolution of solids inventory in the bubbling fluidized bed (inset).}
    \label{bubbleAxiales}
\end{figure}

Fig. \ref{bubbleAxiales} presents the time-averaged axial distribution of solid volume fraction, with a comparison to the numerical solutions available in literature \citep{passalacqua2011ImplementationIterativeSolution,parmentier2008numerical,venier2016numerical}, and an accompanying demonstration of the time evolution of solids inventory (inset). The six implementation methods in this work yield equivalent results for axial solids concentration and solids inventory, which however differ significantly to the available results in literature. The difference could be explained as follow: According to the fact that the initial conditions in all numerical solutions are identical, specifically an initial bed height of 0.2 m, an initial solid volume fraction of 0.58 and the bed height is 1 m, the mean solid volume fraction in the bed ought to be 0.2 $\times$ 0.58/1 = 0.116, which corresponds favorably with our simulation results reported in Fig.\ref{bubbleAxiales}. The numerical outputs from \citet{parmentier2008numerical}, \citet{passalacqua2011ImplementationIterativeSolution} and \citet{venier2016numerical} reveal a time-averaged mean solid volume fraction of 0.1141, 0.0976 and 0.0961 respectively, which are computed through the integral calculation of their digitalized axial solids concentration results and therefore should contain minor errors. The departure from the initial mean solid fraction in the studies of \citet{passalacqua2011ImplementationIterativeSolution} and \citet{venier2016numerical} suggests that their simulations do not hold the critical property of the mass conservation of particles in the bed. On the other hand, the study of \citet{parmentier2008numerical} follows the mass conservation of particles. The observed difference between present simulation and their simulation can be attributed to the difference of used drag model. The drag model of \citet{gidaspow1994multiphase} is used in present study, which is same as in \citet{passalacqua2011ImplementationIterativeSolution} and \citet{venier2016numerical}. In contrast, the drag model of \citet{1966Mechanics} is used in \citet{parmentier2008numerical}.
This is also consistent with the study of \citet{venier2016numerical} and \citet{loha2012assessment} regarding the impact of various drag force models in the current situation.

\begin{figure}[!htb]
    \centering
    \includegraphics[scale=0.45]{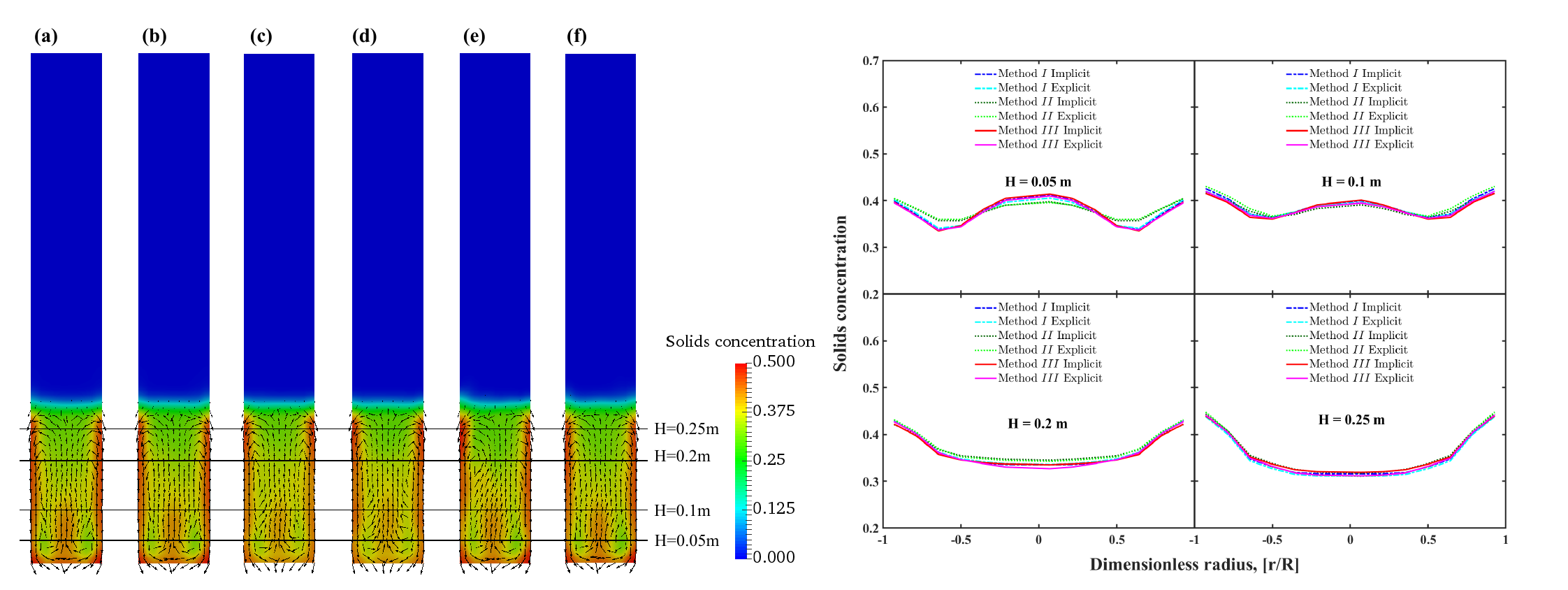}
    \caption{Time-averaged contours of solids concentration with particle velocity vector using (a): Method $\uppercase\expandafter{\romannumeral1}$ Implicit, (b): Method $\uppercase\expandafter{\romannumeral1}$ Explicit, (c): Method $\uppercase\expandafter{\romannumeral2}$ Implicit, (d): Method $\uppercase\expandafter{\romannumeral2}$ Explicit, (e): Method $\uppercase\expandafter{\romannumeral3}$ Implicit, (f): Method $\uppercase\expandafter{\romannumeral3}$ Explicit. Accompanying with the time-averaged cross-sectional radial profile of solids concentration at four different heights in the fluidized bed for six methods.}
    \label{bubble_radial_vector}
\end{figure}

Fig. \ref{bubble_radial_vector} presents the contours that highlight the time-averaged solid volume fraction, along with the particle velocity vector profile and the time-averaged cross-sectional radial profile of solids concentration at four different heights in the fluidized bed, the corresponding heights of the statistics have been identified. It can be illustrated that the detailed bed hydrodynamics remain consistent across the six implementation methods. From the simulation animations, it can be observed that small gas bubbles are continuously generated at the bottom of the bed, gradually coalescing as they ascend and eventually disintegrating at the top. The vector of particle velocities shows that particles ascend at the center region and descend along the wall, which is a typical solid circulation pattern determined by the motion of bubbles \citep{werther1973localII,werther1973local}.
All methods were able to effectively capture the non-uniform character of particles. At the lower part of the bed, Method $\uppercase\expandafter{\romannumeral2}$ Implicit and Explicit obtained consistent time-averaged radial solids concentrations, displaying a more evenly distributed profile than other methods. There is a noticeable trend of decreasing these differences as the fluidized bed height increases.

\subsubsection{Underlying mechanism}
From present analysis and the simulation cases presented above, it is apparent that the six implementation methods of the granular pressure gradient term do not result in significant differences in circumstances as bubbling fluidized beds. The absence of the effect of $\frac{\partial p_s}{\partial \theta_s} \nabla \theta_s$ on $\nabla p_s$ does not have a substantial impact on the accuracy of the outcomes.
Therefore, numerous prior results utilizing the standard solver in OpenFOAM$^{\circledR}$ using Method $\uppercase\expandafter{\romannumeral1}$ for computing $\nabla p_s$ to predict bubbling fluidized beds remain compelling. The observed differences between the simulation results using different software by different authors should be attributed to other issues, such as different constitutive relations, different wall boundary conditions, different input parameters and different algorithms.

\begin{figure}[!htb]
    \centering
    \includegraphics[scale=1.1]{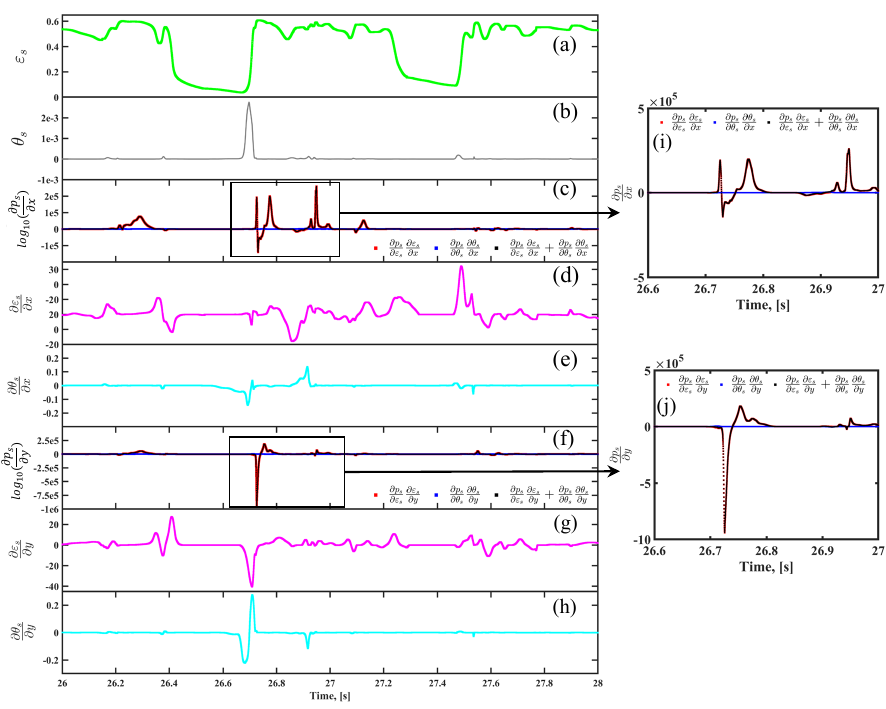}
    \caption{Probed time evolution of (a): solids concentration; (b): granular temperature; terms corresponding to the granular pressure gradient in (c): x-component and (f): y-component with their localized enlargement (i) and (j), respectively; gradient of solids concentration in (d): x-component and (g): y-component, respectively; gradient of granular temperature in (e): x-component and (h): y-component, respectively.}
    \label{figure::bubble_probe}
\end{figure}

In order to explore the underlying mechanism of the minor effect of different implementation methods, probes are placed in the bed. Fig. \ref{figure::bubble_probe} exemplifies the time evolution of the three terms related to the granular pressure gradient that are obtained from the case of Method $\uppercase\expandafter{\romannumeral3}$ Implicit: $\frac{\partial p_s}{\partial \varepsilon_s} \nabla \varepsilon_s$, $\frac{\partial p_s}{\partial \theta_s} \nabla \theta_s$, and their summation $\frac{\partial p_s}{\partial \varepsilon_s} \nabla \varepsilon_s+\frac{\partial p_s}{\partial \theta_s} \nabla \theta_s$ of x-component and  y-component separately, which are obtained from a probe placed in the bubbling fluidized bed $\uppercase\expandafter{\romannumeral1}$ (0.1m from the bed center and 0.4m high).
The solids concentration exhibits significant fluctuations over time (Fig. \ref{figure::bubble_probe}a), suggesting alternative passing of emulsion phase and bubble phase. This phenomenon reflects the temporal bubble movement and fluidization development in the bed. By contrast, the fluctuations and values in granular temperature are significantly smaller (Fig. \ref{figure::bubble_probe}b), with pulsations happening exclusively during sudden variations in solids concentration (the boundaries between the bubble phase and the emulsion phase).
Moreover, it is clear from Fig. \ref{figure::bubble_probe} a, b, d, e, g, h that large gradients of solid volume fraction and granular temperature appear at the boundaries between bubble phase and emulsion phase, which can be easily understood since the solid volume fraction and the granular temperature within the bubble phase and the emulsion phase vary a little. Similarly, the gradient in granular temperature is two orders of magnitude smaller than the gradient in solids concentration.
In general, the three terms linked with $\nabla p_s$ show small values. When pulsations are extensive, the value of $\frac{\partial p_s}{\partial \varepsilon_s} \nabla \varepsilon_s$ term is considerably higher than the value of $\frac{\partial p_s}{\partial \theta_s} \nabla \theta_s$ term by orders of magnitude. Consequently, the sum of the two terms is practically tantamount to the $\frac{\partial p_s}{\partial \varepsilon_s} \nabla \varepsilon_s$ term, and the $\frac{\partial p_s}{\partial \theta_s} \nabla \theta_s$ term serves as a negligible correction to $\nabla p_s$. Hence, it explains the fact that using different implementation methods of $\nabla p_s$ under the bubbling fluidization regime doesn't result in notable disparities in simulation outcomes. It also explains why many previous simulations of gas-solid bubbling fluidization that neglect the contribution of $\frac{\partial p_s}{\partial \theta_s} \nabla \theta_s$ can still obtain a reasonable result.

\subsection{Circulating fluidized bed riser}
\subsubsection{Simulation setup}

\begin{figure}[!htb]
    \centering
        \includegraphics[scale=0.5]{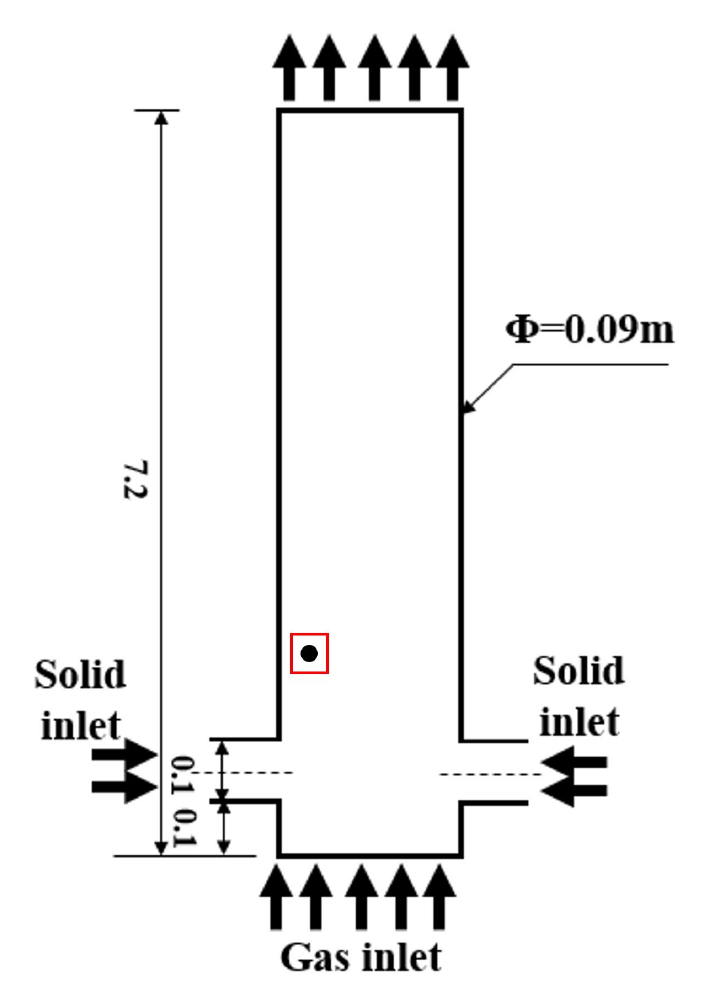}
    \caption{The not-at-scale schematic of the circulating fluidized bed riser and the probe location.}
    \label{figure::probe1}
\end{figure}

A two-dimensional circulating fluidized bed riser is simulated in this study. The geometry of the riser is shown in Fig.\ref{figure::probe1}, where the location of the probe that is used in mechanism analysis is also shown. The gas velocity is specified at the bottom inlet, while the solids inlets are located at the two sides of the riser at the height of 0.1-0.2 m. The outlet is located at the top. Such symmetric inlet and outlet arrangements are more likely to predict appropriate flow patterns and enable quicker grid convergence than their asymmetric counterparts, as shown by the research of \citet{LI2014170}. The parameters used in the simulation are shown in Table. \ref{tab:conf}, which is the same as a previous study \citep{wang2008eulerian}. The boundary conditions are shown in Table. \ref{tab:boundary}, the numerical configurations are shown in Table. \ref{tab:scheme}, and the under-relaxation factors are shown in Table. \ref{tab:relax}. The simulations are performed as transient and after the initialization and development of flow, time-averaged data are then collected, with the fixed time step of 5$\times 10^{-5}$ s. All time steps converged to steady level of residuals. Note that in order to consider the effects of particle clustering structures, an EMMS drag model \citep{lu2009searching} is used in the simulations, details of the drag correlation for each simulations cases however are not reported.

\begin{longtable}{lc}
    \caption{Summary of simulation parameters \label{tab:conf}}
    \\\hline
    \textbf{Properties} & \textbf{Value} \\
    \hline
    Riser diameter $D(m)$                   & 0.09 \\
    Riser height $H(m)$                     & 7.2 \\
    Particle diameter $d_p(\mu m)$              & 100 \\
    Particle density $\rho_s(kg/m^3)$       & 2650 \\
    Gas density $\rho_g(kg/m^3)$            & 1.2 \\
    Gas viscosity $\mu_g(kg/m\cdot s)$      & $1.8\times10^{-5}$ \\
    Superficial gas velocity $U_g(m/s)$     & 4.0\\
    Solid mass flux $G_s(kg/m^2s)$      & 138.5 \quad 150.1 \\
    restitution coefficient             & 0.95 \\
    Initial average voidage in bed $\varepsilon_{s,bed}$      & 0.055 \\
    Grid number (mm$\times$ mm)         & $4.5\times20$ \\
    \hline
\end{longtable}

\begin{longtable}{cccccc}
    \caption{Summary of boundary condition setup, the boundary condition and names of the models are presented as they appear in the software. \label{tab:boundary}}
    \\\hline
    \textbf{Variables} & \textbf{gas\_inlet} & \textbf{solid\_inlet\_left} & \textbf{solid\_inlet\_right} & \textbf{outlet} & \textbf{walls}\\
    \hline
    $\varepsilon_s$ & 0 & 0.2 & 0.2 & zeroGradient & zeroGradient  \\
    $\varepsilon_g$ & 1 & 0.8 & 0.8 & zeroGradient & zeroGradient  \\
    ${\bf u_s}$ & 0 & 0.1176\quad0.1274 & 0.1176\quad0.1274 & pressureInletOutletVelocity & slip  \\
    ${\bf u_g}$ & 4.0 & 0 & 0 & pressureInletOutletVelocity & noslip  \\
    $p$ & zeroGradient & zeroGradient & zeroGradient & 101325Pa & zeroGradient  \\
    $\theta_s$ & 0 & 0.01 & 0.01 & zeroGradient & zeroGradient  \\
    \hline
\end{longtable}

\begin{longtable}{lc}
    \caption{Summary of numerical schemes: $\psi$ denotes a generic variable. The numerical scheme and names of the models are presented as they appear in the software. \label{tab:scheme}}
    \\\hline
    \textbf{Term} & \textbf{Configuration} \\
    \hline
    $\partial / \partial t$ & Euler implicit \\
    $\nabla \psi$ & Gauss linear \\
    $\nabla \cdot \psi$ & Gauss SuperBee \\
    $\nabla \cdot (\nabla \psi)$ & Gauss linear corrected \\
    $\nabla^{\perp}$ & Corrected \\
    $(\psi)_f$ & Gauss linear \\
    \hline
\end{longtable}

\begin{longtable}{lc}
    \caption{Summary of under-relaxation factors \label{tab:relax}}
    \\\hline
    \textbf{Variable} & \textbf{Under-relaxation value} \\
    \hline
    $p$ & 0.5  \\
    $\bm u_k$      & 0.7  \\
    $\varepsilon_k$      & 0.2 \\
    $\theta_s$   & 0.2 \\
    \hline
\end{longtable}

        \subsubsection{Results and discussion}
\begin{figure}[!htb]
    \centering
    \includegraphics[scale=1.1]{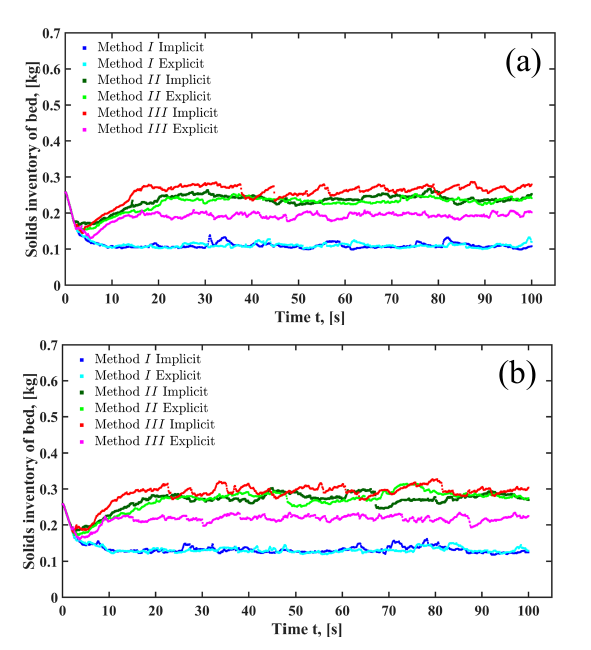}
    \caption{Time evolution of the solids inventory in the riser, (a): $Gs=138.5kg/m^2s$; (b): $Gs=150.1kg/m^2s$.}
    \label{solidinventory}
\end{figure}

The time evolution of solids inventory for two solid mass fluxes are presented in Fig. \ref{solidinventory}.
It is obvious that Method $\uppercase\expandafter{\romannumeral1}$ Implicit and Explicit exhibit a high degree of similarity and own considerably lower solids inventory values compared to other methods.
For Method $\uppercase\expandafter{\romannumeral2}$, the discrepancies between Implicit and Explicit are also trivial.
Apparent discrepancies arise between the implicit and explicit solutions of Method $\uppercase\expandafter{\romannumeral3}$ in both cases, suggesting a paradoxical conclusion that diverse numerical treatments to the gradient of the granular pressure term in the solid-phase continuity equation impacts the ultimate flow dynamics.
According to the figure, it is reasonable to use the data between 70 s to 100 s to calculate the time-averaged statistics.

\begin{figure}[!htb]
    \centering
    \includegraphics[scale=0.6]{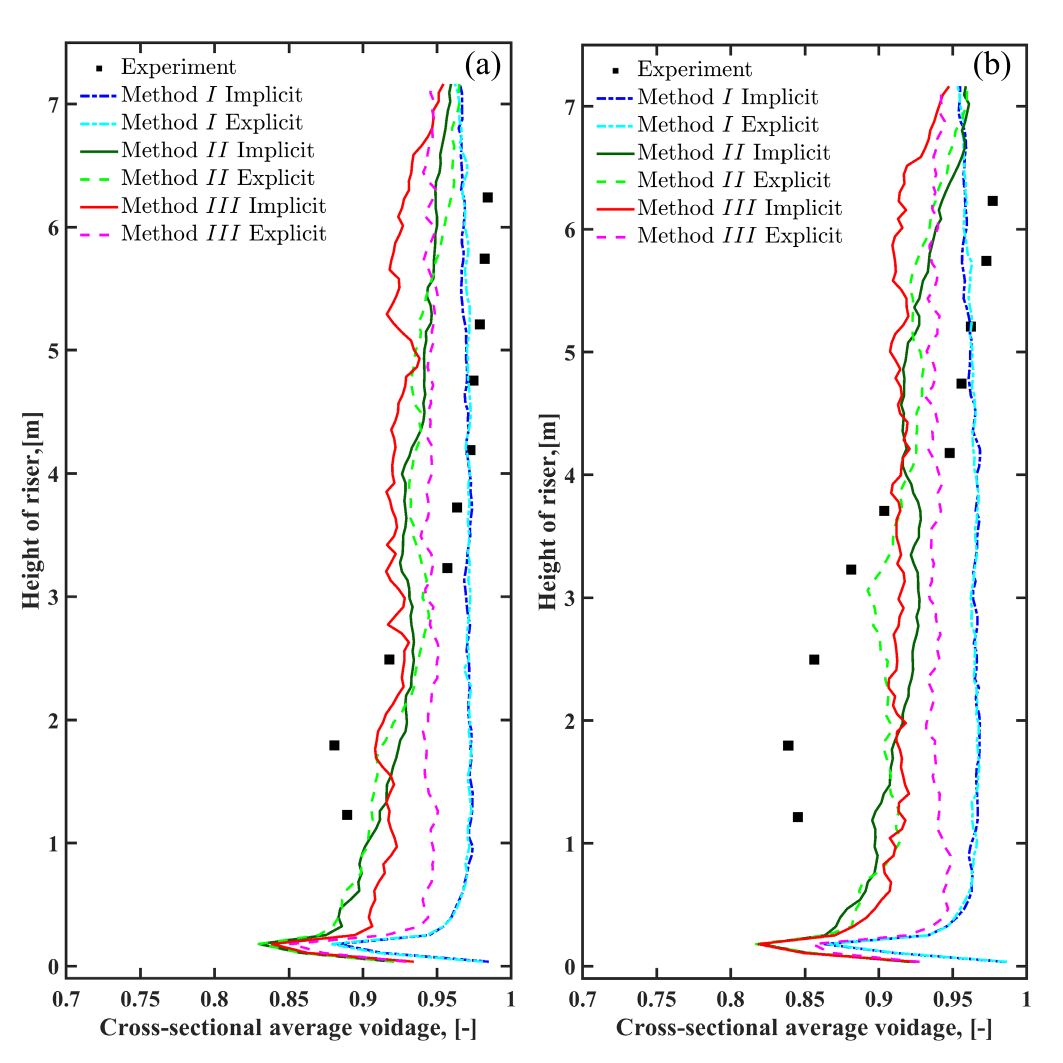}
    \caption{Time averaged axial profiles of cross-sectional averaged solids concentration in the circulating fluidized bed case of (a): $Gs=138.5kg/m^2s$; (b): $Gs=150.1kg/m^2s$.}
    \label{Axiales}
\end{figure}

Fig. \ref{Axiales} compares our time averaged simulation results for divergent cases at given solid mass fluxes ($Gs$) with the experimentally measured axial voidage profiles. Disparities between our simulation results and experimentally measured data can be found in both cases. It appears that Method $\uppercase\expandafter{\romannumeral1}$ and $\uppercase\expandafter{\romannumeral3}$ fail to adequately capture the simultaneous presence of a dilute region in the upper riser and a dense region in the lower section. It however must be stressed here that the comparison between numerical simulations and experimental data critically depends many other issues, such as 2D simulations vs 3D experiments, riser-only simulations vs full-loop experiments, monodisperse particles in simulations vs polydisperse particles in experiments. Therefore, the experimental data here is only for reference, the focal point should be given to the differences arisen from the different treatments of the granular pressure gradient term.

The trend in axial voidage distribution is consistent with the solids inventory. The performance of Implicit and Explicit for Method $\uppercase\expandafter{\romannumeral1}$ and $\uppercase\expandafter{\romannumeral2}$ are almost consistent with each other in the axial behavior throughout the riser height while considerable difference exist in the outcomes of Methods $\uppercase\expandafter{\romannumeral3}$ Implicit and $\uppercase\expandafter{\romannumeral3}$ Explicit.
According to the used KTGF, Method $\uppercase\expandafter{\romannumeral2}$ and $\uppercase\expandafter{\romannumeral3}$ are physically same, since the granular pressure depends on solids concentration and granular temperature. Therefore, the calculation approach for $\nabla p_s$ using the gradient difference scheme in Method $\uppercase\expandafter{\romannumeral2}$ should be equivalent to using the partial derivative summation in Method $\uppercase\expandafter{\romannumeral3}$, if there is no any numerical error. Clearly, the observed differences of simulation results obtained from Method $\uppercase\expandafter{\romannumeral2}$ and $\uppercase\expandafter{\romannumeral3}$ are caused by numerical issues.
A detailed analysis of the mechanisms involved will be presented in section 5.2.3.
Finally, it can be concluded that accurate results in computational fluid dynamics simulations of such two-phase flow require both appropriate implementation approaches and numerical treatment of granular pressure gradients in the solid phase continuity equation.

\begin{figure}[!htb]
    \centering
    \includegraphics[scale=0.5]{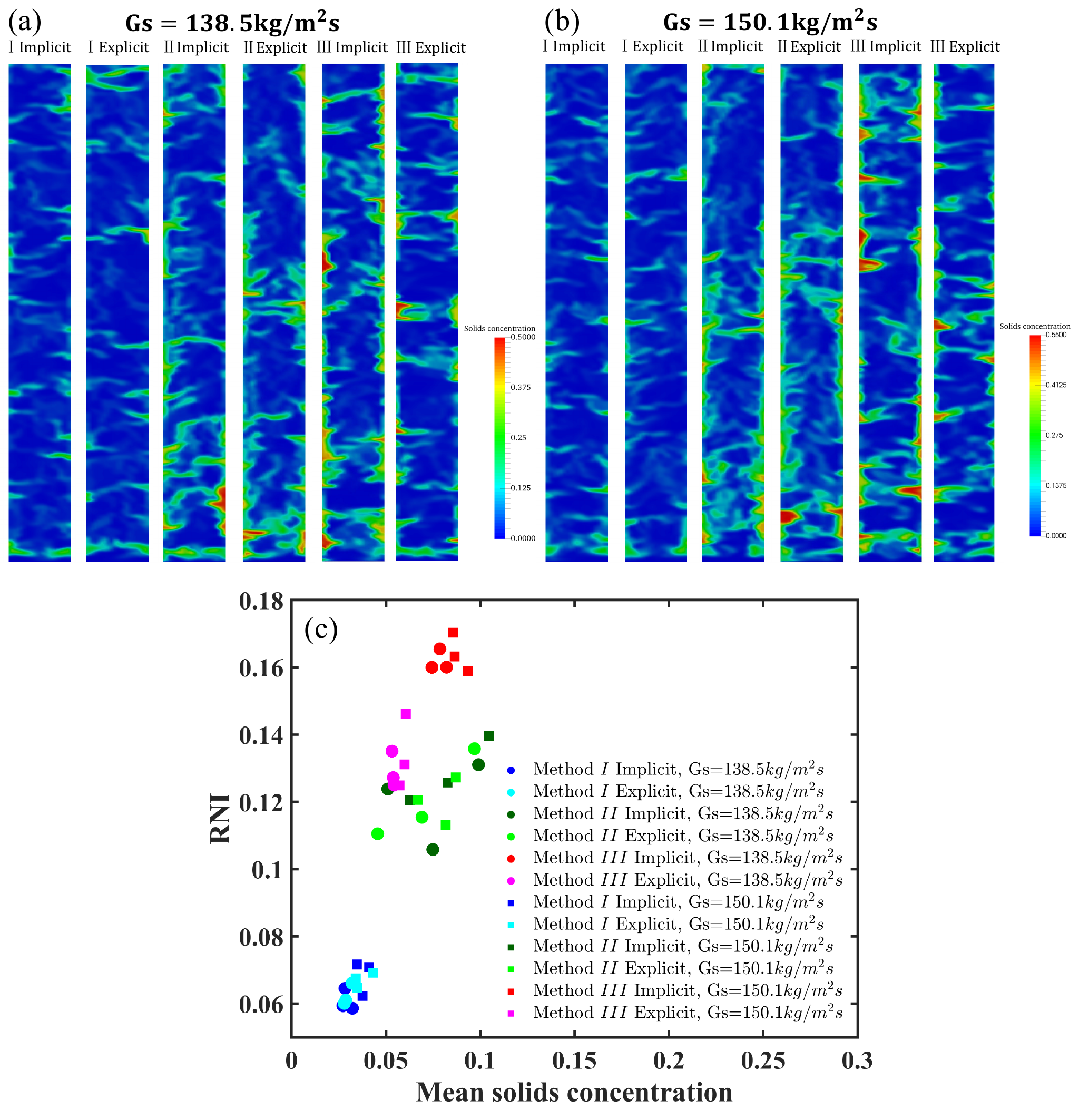}
    \caption{(a), (b): Snapshots at 100 s of various $Gs$ for different $\nabla p_s$ calculation methods and (c): time-averaged radial non-uniformity index (RNI) as a function of cross-sectional solids concentration at different heights of 1 m, 4 m, and 6 m in the riser.}
    \label{RNI}
\end{figure}

Fig. \ref{RNI} (a), (b) present the snapshots at 100 s of various solids fluxes ($Gs$) for different $\nabla p_s$ calculation methods. It is evident from the snapshots that Method $\uppercase\expandafter{\romannumeral2}$ and Method $\uppercase\expandafter{\romannumeral3}$ are able to capture more pronounced mesoscale structures compared to Method $\uppercase\expandafter{\romannumeral1}$.
To quantitatively compare the radial heterogeneity at different heights, \citet{zhu2001radial} defines the standard deviations of radial solids $\sigma(\varepsilon_s)$,
\begin{equation}
\sigma (\varepsilon_s) = \frac{1}{N}\sqrt{\sum_{i=1}^{N}[\varepsilon_s(R_i)-\varepsilon_{s,ave}]^2}
\end{equation}
where $\varepsilon_s(R_i)$ is the local average solids concentration at radius $R_i$, $N$ is the total number of radial data, and $\varepsilon_{s,ave}$ is the average value of $\varepsilon_s$:
\begin{equation}
\varepsilon_{s,ave} = \frac{1}{N}\sum_{i=1}^{N}\varepsilon_s(R_i)
\end{equation}
According to $\sigma(\varepsilon_s)$, the radial non-uniformity index (RNI) can be calculated, which is proven to be an efficient and straightforward method for describing the degree of solid volume fraction radial variation in fluidized beds. RNI is defined as follow by \citet{zhu2001radial}:
\begin{equation}
RNI(\varepsilon_s) = \frac{\sigma(\varepsilon_s)}{\sigma_{max}(\varepsilon_s)}
\end{equation}
where $\sigma_{max}(\varepsilon)$ represents the highest achievable level for standard deviation of $\varepsilon_s$ where solid particles are fully segregated with $\varepsilon_s=0$ at the core, and $\varepsilon_s=\varepsilon_{s,mf}$ at the annulus,
\begin{equation}
\sigma_{max}(\varepsilon_s)=\sqrt{ f( \varepsilon_{s,mf} - \varepsilon_{s,ave} )^2 + (1-f)(\varepsilon_{s,min} - \varepsilon_{s,ave})^2 }
\end{equation}
where $f=\varepsilon_{s,ave}/\varepsilon_{s,mf}$ is the area fraction of the annulus, then we can get:
\begin{equation}
RNI(\varepsilon_s)=\frac{\sigma(\varepsilon_s)}{\sigma_{max}(\varepsilon_s)}=\frac{\sigma(\varepsilon_s)}{\varepsilon_{s,ave}(\varepsilon_{s,mf}-\varepsilon_{s,ave})}
\end{equation}

Fig. \ref{RNI} (c) presents the time averaged RNI($\varepsilon_s$) as a function of cross-sectional solids concentration at different heights of 1 m, 4 m and 6 m in the riser. It is commonly accepted that $\varepsilon_{s,ave}$ increases with solids fluxes ($Gs$). As $\varepsilon_{s,ave}$ increases, RNI($\varepsilon_s$) also increases when the mean solid concentration is smaller than about 0.25. This aligns with the observation of \citet{zhu2001radial} and \citet{wang2008eulerian}. Moreover, Methods $\uppercase\expandafter{\romannumeral2}$ and $\uppercase\expandafter{\romannumeral3}$ exhibit systematically larger non-uniformity than Method $\uppercase\expandafter{\romannumeral1}$, both of which can capture the mesoscale structures in the riser properly.

\subsubsection{Underlying mechanism}
\begin{figure}[!htb]
    \centering
    \includegraphics[scale=0.65]{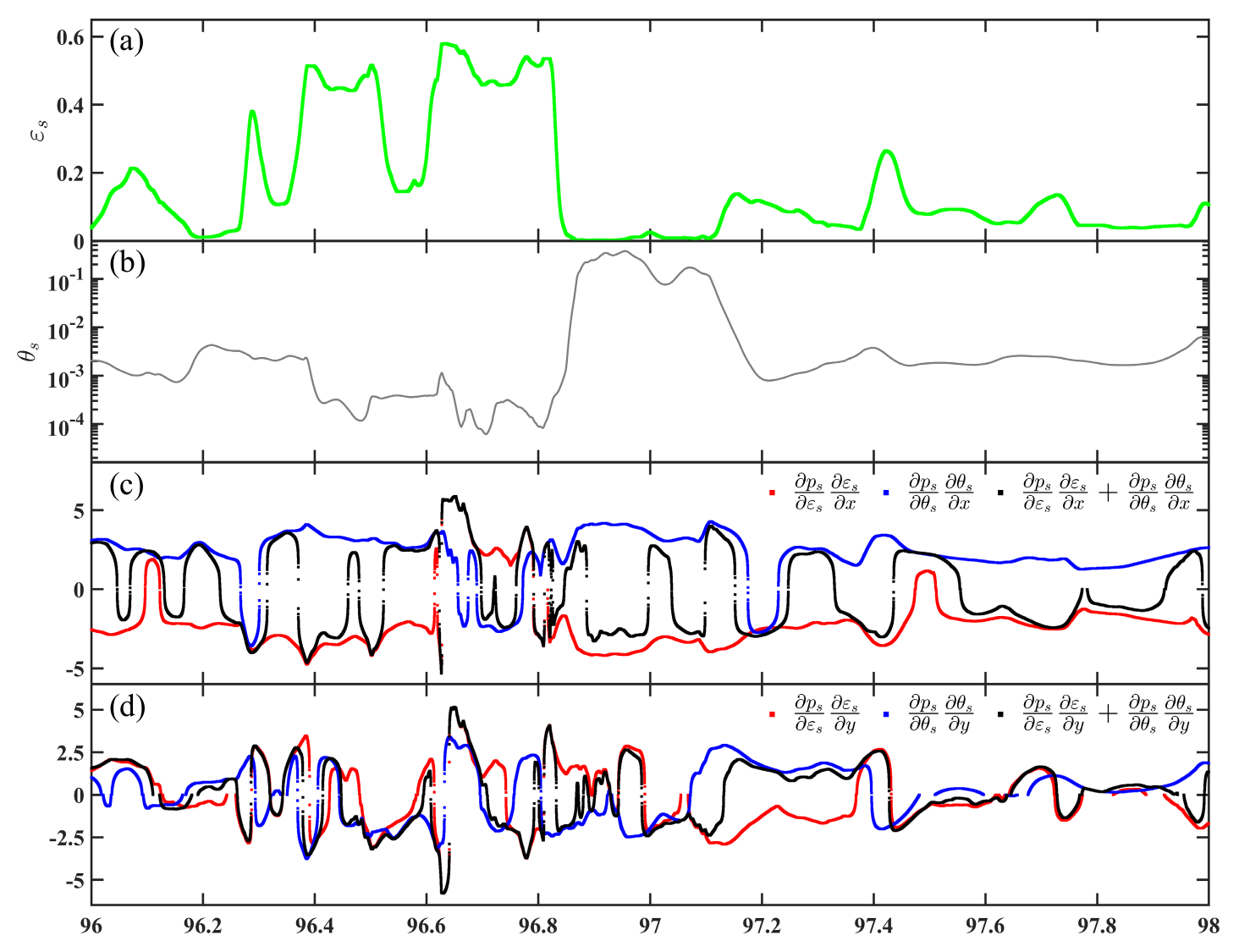}
    \caption{Probed time evolution in Method $\uppercase\expandafter{\romannumeral3}$ of (a): solids concentration; (b): granular temperature; terms corresponding to the granular pressure gradient in (c): x-component and (d): y-component, respectively.}
    \label{figure::riser_probe}
\end{figure}
\begin{figure}[!htb]
    \centering
    \includegraphics[scale=0.55]{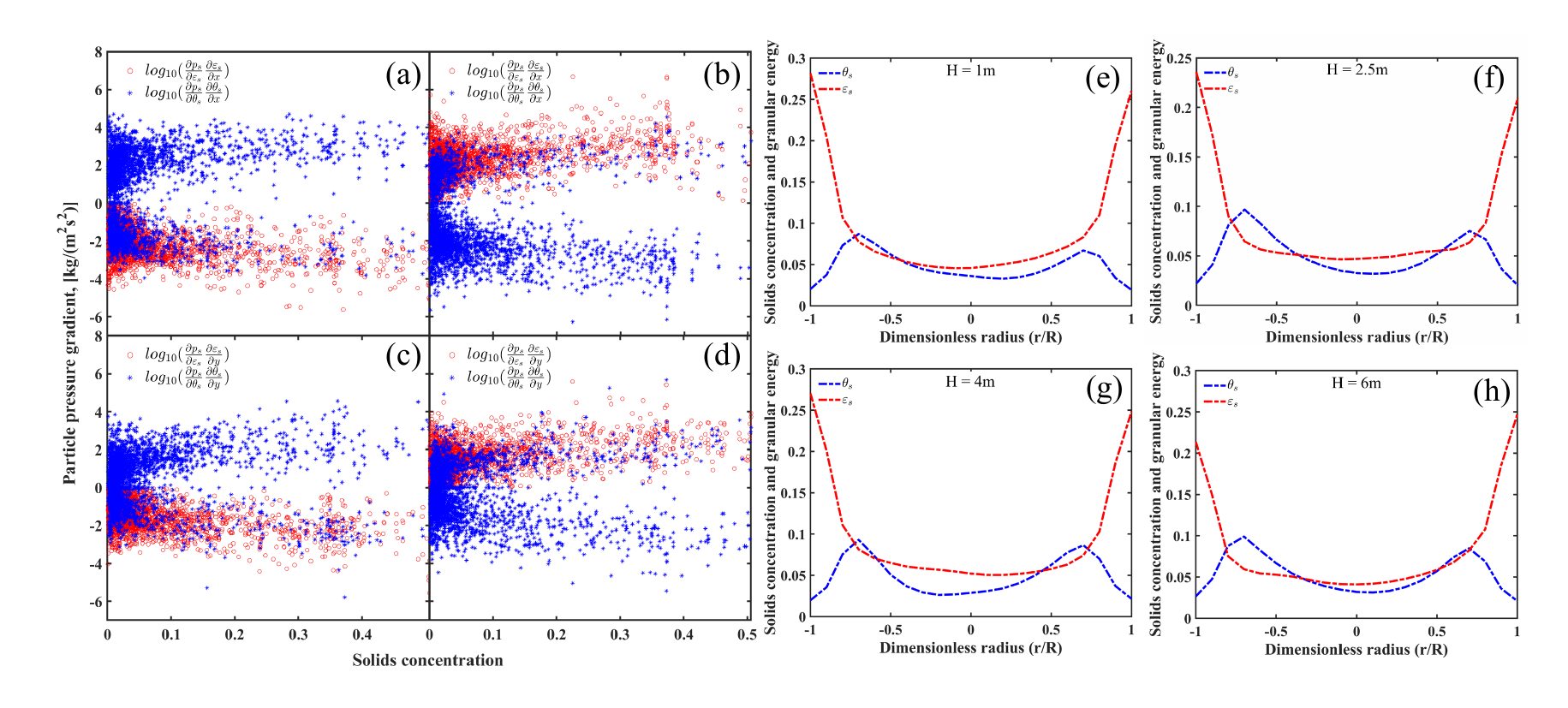}
    \caption{Instantaneous spatial distribution of $\frac{\partial p_s}{\partial \varepsilon_s}\nabla \varepsilon_s$ and $\frac{\partial p_s}{\partial \theta_s}\nabla \theta_s$ as a function of solids concentration in the riser for Method $\uppercase\expandafter{\romannumeral3}$ Implicit, where (a): $\frac{\partial p_s}{\partial \varepsilon_s}\frac{\partial \varepsilon_s}{\partial x}<0$, (b): $\frac{\partial p_s}{\partial \varepsilon_s}\frac{\partial \varepsilon_s}{\partial x}>0$, (c): $\frac{\partial p_s}{\partial \varepsilon_s}\frac{\partial \varepsilon_s}{\partial y}<0$, (d): $\frac{\partial p_s}{\partial \varepsilon_s}\frac{\partial \varepsilon_s}{\partial y}>0$. With time-averaged radial distributions of $\varepsilon_s$ and $\theta_s$ at the height of (e): 1m, (f): 2.5m, (g): 4m, (h): 6m}
    \label{figure::distributions}
\end{figure}
\begin{figure}[!htb]
    \centering
    \includegraphics[scale=0.5]{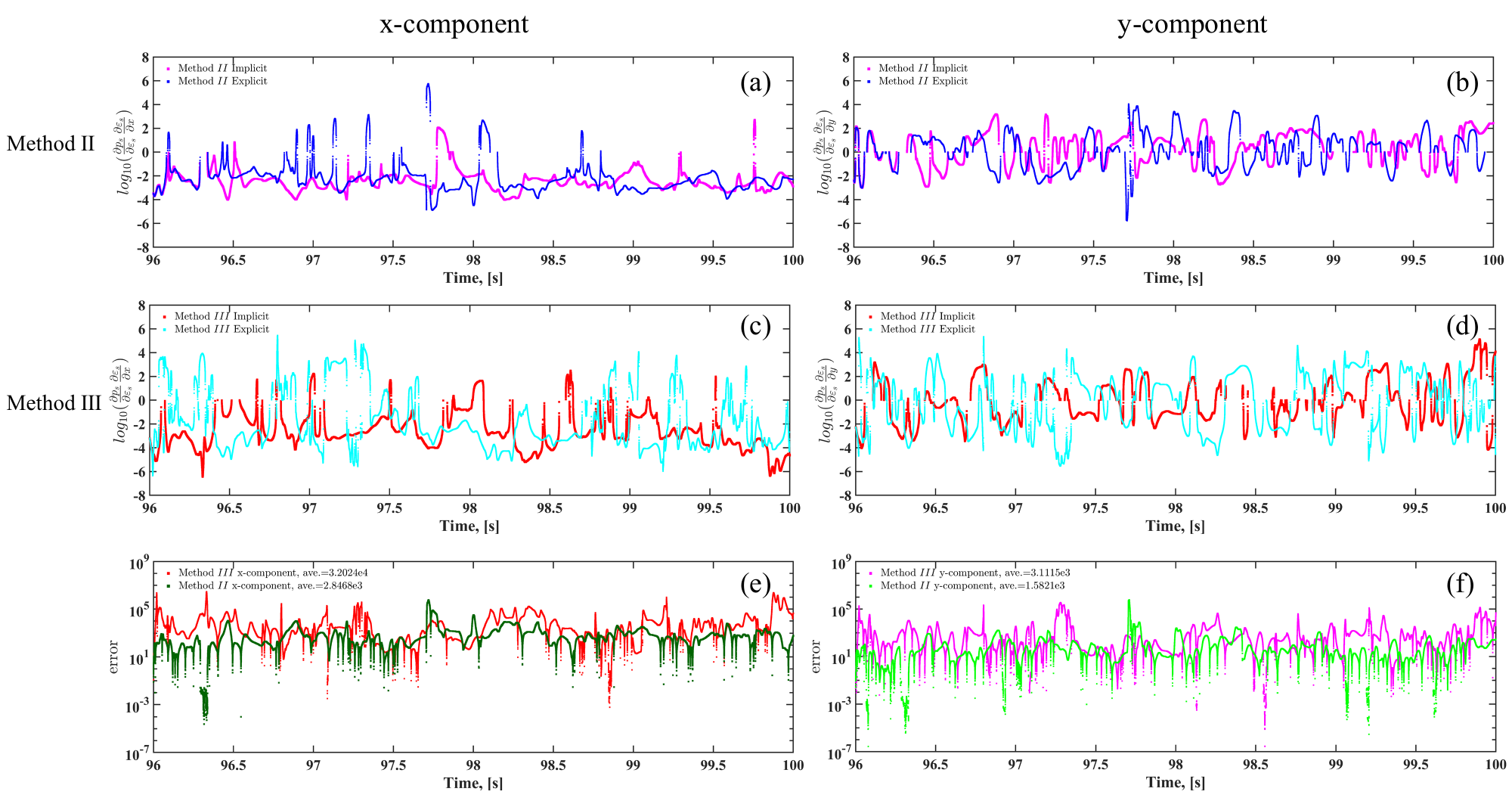}
    \caption{Comparison of implicit and explicit treated granular pressure gradients with respect to solid volume fraction $\frac{\partial p_s}{\partial \varepsilon_s} \nabla \varepsilon_s$ in (a): x-component, Method $\uppercase\expandafter{\romannumeral2}$; (b): y-component, Method $\uppercase\expandafter{\romannumeral2}$; (c): x-component, Method $\uppercase\expandafter{\romannumeral3}$; (d): y-component, Method $\uppercase\expandafter{\romannumeral2}$ and comparisons for error between  Method $\uppercase\expandafter{\romannumeral2}$ and $\uppercase\expandafter{\romannumeral3}$ in (e): x-component; (f): y-component, where error = $|\left(\frac{\partial p_s}{\partial \varepsilon_s} \nabla \varepsilon_s \right)_{Implicit} - \left(\frac{\partial p_s}{\partial \varepsilon_s} \nabla \varepsilon_s \right)_{Explicit}|$.}
    \label{figure::riser_probe_ImEx}
\end{figure}

In order to explore the underlying mechanism of the observations using different implementation methods, probes are placed in the riser as shown in Fig. \ref{figure::probe1} for the case of $Gs=150.1kg/m^2s$. The results of $Gs=138.5kg/m^2s$ are qualitatively same, therefore, they are not shown here. Fig. \ref{figure::riser_probe} exemplifies the time evolution of the three terms related to the granular pressure gradient that are obtained from the case of Method $\uppercase\expandafter{\romannumeral3}$ Implicit: $\frac{\partial p_s}{\partial \varepsilon_s} \nabla \varepsilon_s$, $\frac{\partial p_s}{\partial \theta_s} \nabla \theta_s$, and their summation $\frac{\partial p_s}{\partial \varepsilon_s} \nabla \varepsilon_s+\frac{\partial p_s}{\partial \theta_s} \nabla \theta_s$ of x-component and  y-component respectively, which are obtained from a probe placed in the riser (0.05 m from the wall and 1 m high). When dealing with the data of the granular pressure gradient terms presented in Fig. \ref{figure::riser_probe}-\ref{figure::riser_probe_ImEx}, it is important to note the following:
\begin{itemize}
        \item [(1)]
        Since the obtained data span a large order of magnitude, located in $[-1e6,1e6]$, the positive and negative of the data indicate the direction, and through statistics, it is found that the data between $[-1,1]$ only accounts for a few out of ten thousand, and physically this part of the data has little influence, so the data between $[-1,1]$ were removed, and the remaining parts were taken as the logarithm base 10.
        \item [(2)]
        When $\frac{\partial p_s}{\partial \varepsilon_s} \nabla \varepsilon_s > 1$, the log value is $log_{10}(\frac{\partial p_s}{\partial \varepsilon_s} \nabla \varepsilon_s)$, when $\frac{\partial p_s}{\partial \varepsilon_s} \nabla \varepsilon_s < -1$, the log value is $-log_{10}(-\frac{\partial p_s}{\partial \varepsilon_s} \nabla \varepsilon_s)$. After the processing, the plus and minus signs for the resulting data represent directions, and their absolute value are taken over the logarithmic of base 10. The same processing is done for $\frac{\partial p_s}{\partial \theta_s} \nabla \theta_s$ and $\frac{\partial p_s}{\partial \varepsilon_s} \nabla \varepsilon_s+\frac{\partial p_s}{\partial \theta_s} \nabla \theta_s$.
 \end{itemize}

The way in which the solids concentration (Fig. \ref{figure::riser_probe}a) and granular temperature (Fig. \ref{figure::riser_probe}b) changes over time are qualitatively comparable to that of bubbling fluidized bed (Fig. \ref{figure::bubble_probe}). However, in this case, the value of the granular temperature increased by several orders of magnitude, suggesting that the granular temperature contributes substantially more to the granular pressure.
Besides, in most instances of Fig. \ref{figure::riser_probe} c, d, $\nabla p_s$ show larger values than those in bubbling fluidised beds. More importantly, the term with respect to solids concentration $\frac{\partial p_s}{\partial \varepsilon_s} \nabla \varepsilon_s$ and the term with respect to granular temperature $\frac{\partial p_s}{\partial \theta_s} \nabla \theta_s$ have values of comparable magnitude but with different signs, indicating their contrasting impacts on the formation of clusters, making their sum $\frac{\partial p_s}{\partial \varepsilon_s} \nabla \varepsilon_s+\frac{\partial p_s}{\partial \theta_s} \nabla \theta_s$ lies between them. Since the granular pressure gradient takes the form of a negative term $-\nabla p_s$ in the governing equations, the role of term $\frac{\partial p_s}{\partial \varepsilon_s} \nabla \varepsilon_s$ is to inhibit the formation of clustering structures. While the term $\frac{\partial p_s}{\partial \theta_s} \nabla \theta_s$, on the other hand, promotes particle clustering. Therefore, the discrepancies in results between Method $\uppercase\expandafter{\romannumeral1}$ and Method $\uppercase\expandafter{\romannumeral3}$ can be reasonably illustrated. It is the compensatory term $\frac{\partial p_s}{\partial \theta_s} \nabla \theta_s$ that will reduce the absolute values of the granular pressure gradient and facilitate mesoscale structures.
Consequently, incorporating the calculation methods of granular pressure gradient term that corrected by granular temperature becomes crucial in such regime as circulating fluidization. The difference between Method $\uppercase\expandafter{\romannumeral1}$ and Method $\uppercase\expandafter{\romannumeral2}$ can be qualitatively explained in a same way.

Fig. \ref{figure::distributions} a-d represents the instantaneous spatial distribution of two terms related to the granular pressure gradients as a formation of partial derivative, one is concerning solid volume fraction $\frac{\partial p_s}{\partial \varepsilon_s} \nabla \varepsilon_s$ and the other is concerning granular temperature $\frac{\partial p_s}{\partial \theta_s} \nabla \theta_s$ separately, as a function of solid volume fraction. Data are exported from case of Method $\uppercase\expandafter{\romannumeral3}$ Implicit. The distributions are classified by the directions and components of $\frac{\partial p_s}{\partial \varepsilon_s}\nabla \varepsilon_s$, with its matching distribution of $\frac {\partial p_s} {\partial \theta_s} \nabla \theta_s$.
The distributions of same sign and opposite signs for $\frac{\partial p_s}{\partial \varepsilon_s}\nabla \varepsilon_s$ and $\frac{\partial p_s}{\partial \theta_s}\theta_s$ are comparable at lower solids concentrations. However, as the solids concentration increases, there is a gradual increase in the proportion of both having opposite signs. In the region where solids concentration exceeds some value between 0.05 and 1, the majority of the two exhibits distinct differences, as evidenced by opposite signs. Thus, it can be concluded that the correction of the granular pressure gradient by the granular temperature gradient is more prominent at higher solids concentrations.

Fig. \ref{figure::distributions} e-h represents the time-averaged radial distribution of $\varepsilon_s$ and $\theta_s$ at different heights in the riser. The solid volume fraction near the wall is higher, and it's qualitative trend is opposite to that of the granular temperature. So the gradients of the two is of the opposite direction. In the center region where the solid volume fraction is lower, the qualitative trend of solid volume fraction is the same as that of the granular temperature. So the gradients of the two is of the same direction. It can be inferred that this conclusion holds true at various heights in the riser, but only the critical value of the solid volume fraction for both gradients change from the same direction to opposite direction varies with heights, which located between 0.05 and 1. In summary, the conclusion obtained from Fig. \ref{figure::distributions} e-h are consistent with Fig. \ref{figure::distributions} a-d and Fig. \ref{figure::riser_probe} c, d.

Fig. \ref{figure::riser_probe_ImEx} illustrates the probed time evolution results of granular pressure gradients term $\frac{\partial p_s}{\partial \varepsilon_s} \nabla \varepsilon_s$, comparison between implicit and explicit numerical treatment in the solid phase continuity equation are made. Only the term related to the solid volume fraction is compared because it is the distinguishing factor between the different numerical treatments and the other term $\frac{\partial p_s}{\partial \theta_s} \nabla \theta_s$ is always explicitly treated (Eq. (\ref{equ:continuous1})). Note that the initial state of 96 s is the same for all four cases. First, the implicit and explicit results of Method $\uppercase\expandafter{\romannumeral2}$ are almost identical for the first 0.4 s, approximately. However, they gradually diverge over time. The average absolute deviation in the x-direction is 2.8468e3 and in the y-direction is 1.5821e3 for  the whole 96-100 s.
Next, Method $\uppercase\expandafter{\romannumeral3}$ Implicit and $\uppercase\expandafter{\romannumeral3}$ Explicit quickly produce a larger difference within the first few time steps. The average absolute deviation for Method $\uppercase\expandafter{\romannumeral3}$ in the x-direction is 3.2024e4 and in the y-direction is 3.1115e3, which are significantly larger than that of Method $\uppercase\expandafter{\romannumeral2}$.
Therefore, when solving the solid-phase continuity equation, the effect of $\frac{\partial p_s}{\partial \varepsilon_s} \nabla \varepsilon_s$'s perturbations on $\varepsilon_s$ is more pronounced with $\uppercase\expandafter{\romannumeral3}$ than with $\uppercase\expandafter{\romannumeral2}$. Consequently, the difference between implicit and explicit treatments is more significant for Method $\uppercase\expandafter{\romannumeral3}$ than $\uppercase\expandafter{\romannumeral2}$. In summary, compared to Method $\uppercase\expandafter{\romannumeral2}$, Method $\uppercase\expandafter{\romannumeral3}$ engenders a system where the solid volume fraction varies more sensitively with perturbations in the granular pressure gradient term $\frac{\partial p_s}{\partial \varepsilon_s} \nabla \varepsilon_s$.

\begin{figure}[!htb]
    \centering
    \includegraphics[scale=0.8]{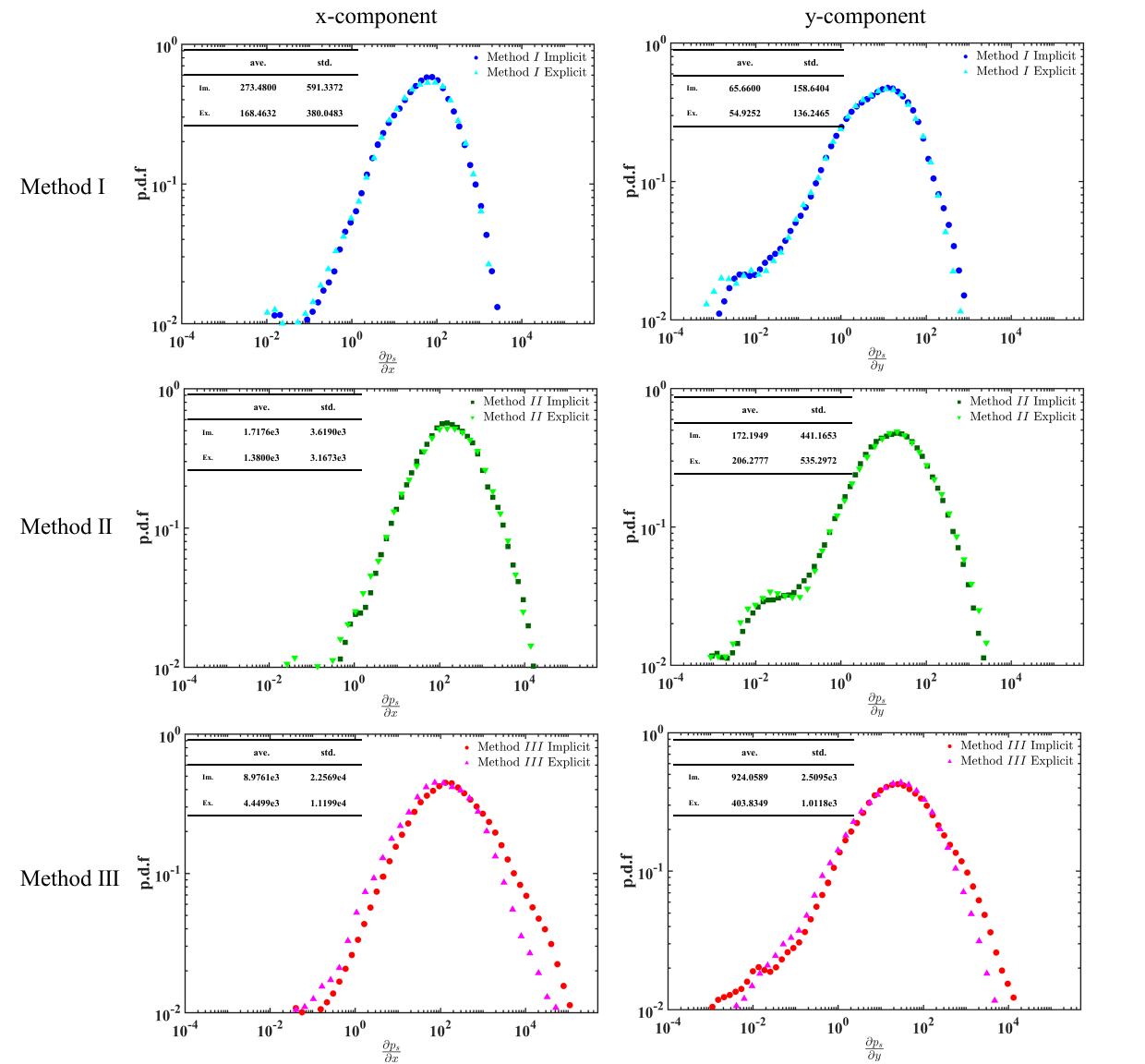}
    \caption{Probability density function comparisons of $|\frac{\partial p_s}{\partial \varepsilon_s} \nabla \varepsilon_s|$ in the riser between implicit and explicit treatment for Method $\uppercase\expandafter{\romannumeral1}$, $\uppercase\expandafter{\romannumeral2}$ and $\uppercase\expandafter{\romannumeral3}$ in x- and y-component.}
    \label{figure::riser_probe_ImEx2}
\end{figure}

To examine the spatial arrangement, Fig. \ref{figure::riser_probe_ImEx2} compares the probability density function (p.d.f) of $|\frac{\partial p_s}{\partial \varepsilon_s} \nabla \varepsilon_s|$ in the riser between implicit and explicit treatment for Method $\uppercase\expandafter{\romannumeral1}$, $\uppercase\expandafter{\romannumeral2}$ and $\uppercase\expandafter{\romannumeral3}$ in x- and y-component respectively. Note that only the data having a probability larger than 0.01 are shown. In this interval, Methods $\uppercase\expandafter{\romannumeral1}$ and $\uppercase\expandafter{\romannumeral2}$ showed overlapping implicit and explicit curves, while Method $\uppercase\expandafter{\romannumeral3}$ showed a mismatch between them. This finding is consistent with the time-averaged results. The accompanying forms list the corresponding mathematical expectations and standard deviations. It is observed that the difference between implicit and explicit increases from Methods $\uppercase\expandafter{\romannumeral1}$ to $\uppercase\expandafter{\romannumeral2}$, and to $\uppercase\expandafter{\romannumeral3}$, particularly for the x-direction, where the difference spans several orders of magnitude.

It is well accepted that in Method $\uppercase\expandafter{\romannumeral1}$, the granular pressure gradient inherently serve as a kind of inter-particle repulsive force, which increases dramatically with increasing gradient of solid volume fraction. Consequently, this effect pushes the particles away from each other and effectively curtails their excessive accumulation. In Method $\uppercase\expandafter{\romannumeral2}$ and $\uppercase\expandafter{\romannumeral3}$, the contribution of granular temperature is compensated in order to physically conform with the KTGF theory. However, due to the typically opposing distributions of solids concentration and granular temperature in the fluidized bed \citep{lu2008numerical,muller2008granular, tartan2004measurement}, the gradient of solids concentration $\nabla \varepsilon_s$ and granular temperature $\nabla \theta_s$ are naturally opposite in direction at regions with higher solids concentration (Fig. \ref{figure::distributions} e-h). The impact of granular temperature weaken or neutralize the effect of $\nabla \varepsilon_s$ on $\nabla p_s$ (Fig. \ref{figure::distributions} a-d, Fig. \ref{figure::riser_probe} c, d). Therefore, the absolute value of granular pressure gradient $\nabla p_s$ is corrected to a smaller value, which is more inclined to the formation of clusters. This leads to distinctions in the axial distribution of solids concentration (Fig. \ref{Axiales}), increase in radial non-uniformity of solids concentration RNI($\varepsilon_s$) (Fig. \ref{RNI}) and significant increases in the overall solids inventory in the riser. In addition, the formation and release of clusters result in fluctuations in the solids inventory curve (Fig. \ref{solidinventory}). Compared to the implicit numerical treatment for $p_s$ in the solid phase continuity equation, the explicit strategy has a delay at each iteration within each time step, and ultimately leading to disparate hydrodynamic behaviors during prolonged periods. This phenomenon is especially noticeable in Method $\uppercase\expandafter{\romannumeral3}$, but has minimal impact on Method $\uppercase\expandafter{\romannumeral1}$ and $\uppercase\expandafter{\romannumeral2}$.

\section{Conclusions}
Three methods for calculating the granular pressure gradient for gas-solid two-fluid systems, which are implemented in open source software OpenFOAM$^{\circledR}$, are investigated.
Moreover, implicit treatment of granular pressure gradient term in the solid phase continuity equation is also realized to address the highly nonlinear dependence on the solid concentration. Results are compared to explicit strategy to investigate the effect of numerical solutions. The conclusions are summarized as follows:
\begin{itemize}
\item[1.]In the bubbling fluidized bed cases, there is no significant difference between the three calculation methods. The influence of missing $\nabla \theta_s$ on $\nabla p_s$ does not remarkably influence the accuracy of results. Further exploration of the underlying mechanism indicates that in the bubbling regime, the granular temperature gradient term has a negligible effect on the granular pressure gradient. Instead, it is the solid volume fraction gradient term that plays a decisive role. Therefore, the outcomes obtained by a large number of previous scholars using OpenFOAM$^{\circledR}$ standard solver, namely Method $\uppercase\expandafter{\romannumeral1}$, to model the bubbling beds are still convincing.
\item[2.]In the case of circulating fluidized beds, there are noticeable impacts among these calculation methods. The reason is that the spatio-temporal hydrodynamic evolve more drastically with intense phase transitions in the circulating fluidized regime compared to bubbling regime. (i) Solving the particle pressure term implicitly or explicitly in the solid-phase continuity equation will produce disparities in the results for Method $\uppercase\expandafter{\romannumeral3}$ but has little or no impact for methods $\uppercase\expandafter{\romannumeral1}$ and $\uppercase\expandafter{\romannumeral2}$. The spatial and temporal distribution of the granular gradient $\frac{\partial p_s}{\partial \varepsilon_s} \nabla \varepsilon_s$ in the riser are also in accordance with the outcomes. It seems that the distribution of solid volume fraction in the system of Method $\uppercase\expandafter{\romannumeral3}$ is more sensitive to perturbations in the granular pressure gradient than the other two methods.  (ii) Various methods for calculating the granular pressure gradient can affect flow results. Method $\uppercase\expandafter{\romannumeral2}$ and $\uppercase\expandafter{\romannumeral3}$, which have shared physical and mathematical nature, differ considerably from method $\uppercase\expandafter{\romannumeral1}$, mainly in the capture of non-uniform structures. Additionally, in the region of higher solids concentration. The granular pressure gradient term related to the granular temperature $\frac{\partial p_s}{\partial \theta_s} \nabla \theta_s$ and the term related to the solid volume fraction $\frac{\partial p_s}{\partial \varepsilon_s} \nabla \varepsilon_s$ have mutually exclusive roles in the formation of mesoscale structures. The supplementary influence regard to the granular temperature will correct the granular pressure gradient to a smaller value, making it easier to promote the formation of non-uniform structures.
\end{itemize}

To sum up, the newly implemented Methods $\uppercase\expandafter{\romannumeral2}$ and $\uppercase\expandafter{\romannumeral3}$ in  OpenFOAM$^{\circledR}$ for calculating the granular pressure gradient term, which complement the contribution of the granular temperature, are more consistent with the hypothesis of KTGF in physical nature. The simulation outcomes are more reasonable compared to primitive Methods $\uppercase\expandafter{\romannumeral1}$. Besides, implicit treatment for the granular pressure gradient is preferable to explicit treatment.
\section*{Declaration of Competing Interest}
The authors declare that they have no known competing financial interests or personal relationships that could have appeared to influence the work reported in this paper.

\section*{Acknowledgement}
This study is financially supported by the National Natural Science Foundation of China (11988102, 21978295, 22311530057), the Strategic Priority Research Program of the Chinese Academy of Sciences (XDA29040200), and the Young Elite Scientists Sponsorship Program by CAST (2022QNRC001).

\end{spacing}
\end{document}